\documentclass{article}
\usepackage[utf8]{inputenc}
\usepackage{dirtytalk}
\usepackage{style}
\newcommand*\samethanks[1][\value{footnote}]{\footnotemark[#1]}

\title{
Linear Aggregation in Tree-based Estimators}
\author{%
\name S\"oren R. K\"unzel\thanks{\hspace{.05in}These authors contributed equally to this work.}
\email srk@berkeley.edu\\
\addr Department of Statistics\\
University of California, Berkeley 
\AND
\name Theo F. Saarinen\samethanks 
\email theo\_s@berkeley.edu\\
\addr 
University of California, Berkeley
\AND
\name Edward W. Liu 
\email edwardliu@berkeley.edu \\
\addr University of California, Berkeley
\AND
\name Jasjeet S. Sekhon\thanks{\hspace{.06in}Saarinen and Sekhon thank ONR grant N00014-17-1-2176.} 
\thanks{\hspace{.06in}The authors would like to thank two anonymous reviewers, the associate editor, and the editor Tyler McCormick for useful comments and suggestions.}
\thanks{\hspace{.06in}Code for replication of results can be found at \url{https://github.com/forestry-labs/RidgePaperReplication}.}
\thanks{\hspace{.06in}The \textbf{Rforestry} package can be found at \url{https://github.com/forestry-labs/Rforestry}.} 
\email jas.sekhon@yale.edu\\
\addr Department of Statistics \& Data Science\\
Yale University
}
\date{\today}
\usepackage{comment}
\usepackage{hyperref}

\usepackage[round]{natbib}
\graphicspath{{figures/001IntroFigure/}{figures/}{figures/Interpret_try2/datasets/GOTV/trees_full/}{figures/2-VaryNSim/}{figures/001_IntroFigure/}}
\usepackage[section]{placeins}
\usepackage{float}
\usepackage{subcaption}
\usepackage{amsmath}
\usepackage{amsfonts}
\usepackage{amssymb}
\usepackage{color}
\usepackage{xspace}
\usepackage{listings}
\usepackage[table]{xcolor}
\usepackage{subfloat}

\usepackage[noend]{algpseudocode}
\usepackage{placeins}
\usepackage{algorithm}
\usepackage{etoolbox}
\usepackage{amssymb}

\makeatletter  
\def\BState{\State\hskip-\ALG@thistlm}
\makeatother
\makeatletter
\AfterEndEnvironment{algorithm}{\let\@algcomment\relax}
\AtEndEnvironment{algorithm}{\kern2pt\hrule\relax\vskip3pt\@algcomment}
\let\@algcomment\relax

\newcommand\algcomment[1]{\def\@algcomment{\footnotesize#1}}
\newcommand{\XiOT}{\begin{bmatrix} X_i^T & 1\end{bmatrix}}
\newcommand{\XiO}{\begin{bmatrix} X_i \\ 1\end{bmatrix}}
\newcommand{\XkOT}{\begin{bmatrix} X_k^T & 1\end{bmatrix}}
\newcommand{\XkO}{\begin{bmatrix} X_k \\ 1\end{bmatrix}}
\newcommand{\E}{\mathbb{E}}
\newcommand{\ranger}{\textbf{\textit{ranger}}\xspace}
\newcommand{\forestry}{\textbf{\textit{Rforestry}}\xspace}
\newcommand{\glmnet}{\textbf{\textit{glmnet}}\xspace}
\newcommand{\caret}{\textbf{\textit{caret}}\xspace}

\newcommand{\dbarts}{\textbf{\textit{dbarts}}\xspace}
\newcommand{\grf}{\textbf{\textit{grf}}\xspace}
\newcommand{\Cubist}{\textbf{\textit{Cubist}}\xspace}
\newcommand{\LRF}{\textit{LRF}\xspace}

\newcommand{\bigO}{\mathcal{O}}
\newcommand{\R}{\mathbb{R}}

\newcommand{\Ybar}{\bar Y}
\newcommand{\RSS}{\mbox{RSS}}
\newcommand{\define}{:=}
\renewcommand{\O}{\mathcal{O}}
\newcommand{\f}{\ell}
\newcommand{\code}{\lstinline[columns=fixed]}
\newcommand{\inputNL}{		\\\hspace*{\algorithmicindent}\hphantom{\textbf{Input: }}~}

\DeclareMathOperator*{\argmin}{arg\,min}

\newtheorem{theorem}{Theorem}

\begin{document}

\maketitle

\begin{abstract}
    Regression trees and their ensemble methods are popular methods for nonparametric regression: they combine strong predictive performance with interpretable estimators. To improve their utility for locally smooth response surfaces, we study regression trees and random forests with linear aggregation functions. We introduce a new algorithm that finds the best axis-aligned split to fit linear aggregation functions on the corresponding nodes, and we offer a quasilinear time implementation. We demonstrate the algorithm's favorable performance on real-world benchmarks and in an extensive simulation study, and we demonstrate its improved interpretability using a large get-out-the-vote experiment.  We provide an open-source software package that implements several tree-based estimators with linear aggregation functions.
\end{abstract}

\section{Introduction}
Classification and Regression Trees (CART) \citep{morgan1963problems, breiman1984classification} have long been used in many domains. Their simple structure makes them interpretable and useful for statistical inference, data mining, and visualizations.
Random Forests (RF) \citep{breiman2001random} and Gradient Boosting Machines (GBM) \citep{friedman2001greedy} build on tree algorithms. They are less interpretable and more laborious to visualize than single trees, but they often perform better in predictive tasks and lead to smoother estimates \citep{buhlmann2002analyzing, svetnik2003random, touw2012data}.

As these tree-based algorithms predict piece-wise constant responses, this leads to (a) weaker performance when the underlying data generating process exhibits smoothness and (b) relatively deep trees that are harder to interpret.

To address these weaknesses, we study regression trees with linear aggregation functions implemented in the leaves. Specifically, we introduce three core changes to the classical CART algorithm:
\begin{enumerate}
    \item 
        Instead of returning for each leaf the mean of the corresponding training observations as in the classical CART algorithm, we fit a ridge regression in each leaf. That is, for a unit with features, $x_{\mbox{\tiny new}}$, that falls in a leaf $S$, the tree prediction is
        \begin{equation} \label{eq:linearRet}
            \hat \mu(x_{\mbox{\tiny new}}) \define x_{\mbox{\tiny new}}^t (X_S^t X_S + \lambda I)^{-1} X_S^t Y_S,
        \end{equation}
        where $Y_S$ is the vector of $y$-values of the training observations that fall in leaf $S$, $X_S$ is the corresponding design matrix for these observations, and $\lambda \in \mathbb{R}^+$ is a regularization parameter.\footnote{In a more elaborate version of the algorithm, the observations in leaf $S$ are a subset of training observations that is disjoint from the training observations used to create the skeleton of the trees. This implements \textit{honesty} \citep{biau2010analysisrf}.}
    \item 
        Crucial to the success of such an aggregation function, we take into account the fact that we fit a regularized linear model on the leaves when constructing the tree by following a greedy strategy that finds the \textit{optimal} (we define what we mean by \textit{optimal} in Section \ref{section:Algorithm}) splitting point at each node. 
        That is, for each candidate split that leads to two child nodes $S_L$ and $S_R$, we take the split such that the total MSE after fitting an $\ell_2$-penalized linear model on each child leaf is minimized.  
        This is difficult, and one of our main contributions is that we found a computationally efficient way to find this optimal splitting point. We discuss in detail how the optimal splitting point is calculated in Section \ref{section:Algorithm}.
    \item 
        Furthermore, we use a cross-validation stopping criteria that determines when to continue splitting as opposed to when to construct a leaf node. After selecting the optimal split, the improvement in $R^2$ that is introduced by the potential split is calculated. If the potential split increases the $R^2$ by less than a predetermined percentage, then the splitting procedure is halted and a leaf node is created. This leads to trees that can create large nodes with smooth aggregation functions on smooth parts of data, while taking much smaller nodes which mimic the performance of standard CART aggregation on separate subsets of the response surface. This adaptivity over varying degrees of smoothness is displayed below in Figure \ref{fig:classicalVSlinearCART} and explored in greater detail in Section \ref{section:simulationstudies}.
\end{enumerate}

These changes improve the \textbf{predictive performance} of such Linear Regression Trees (LRT) substantially, and---as we see in Section \ref{section:simulationstudies}--- our algorithm compares favorably on real-world examples when used in a Linear Random Forests (LRF) ensemble.
Several comments expressed concerns that LRF may not offer improvements over several existing competitors, namely the GBM and RuleFit algorithm. 
In Section \ref{section:simulationstudies}, we see that across a range of setups, LRF has consistently the highest average performance ranking.
We also connect our algorithm with an effective hyperparameter tuning algorithm, and we find that it can behave like both a regularized linear model and a CART/RF estimator depending on the chosen hyperparameters. This is explored in a set of experiments in Section \ref{section:simulationstudies}. 

The linear aggregation functions lead to much shallower trees without losing predictive performance. This improves the \textbf{interpretability} of the trees substantially---a direction which has always been a major motivation for studying regression trees with linear aggregation functions \citep{karalivc1992employing}. 

When the actual underlying problem is smooth, as is the case in many natural phenomena, the method's ability to express that smoothness in an interpretable way is especially important. As Efron notes, it is not a coincidence that Newton’s calculus accompanied Newton’s laws of motion: the underlying physical phenomena itself is an \say{infinitely smooth one in which small changes in cause yield small changes in effect; a world where derivatives of many orders make physical sense} \citep[][646--647]{efron2020prediction}. The use of linear response functions helps researchers discover smooth structures in the data generating process, unlike RF and GBM.\footnote{As with parametric models, basis expansions can be included in the linear response functions to capture interpretable and smooth non-linear functions, with ridge regularization.}

Finally, in Section \ref{section:interpretability}, we apply our algorithm to a causal inference problem. Specifically, we illustrate the advantages for interpretability our algorithm has using a large data set measuring the treatment effect of several treatments on voter turnout \citep{GerberGreenLarimer}.  We adapt the S-Learner with random forests as implemented in causalToolbox \citep{kunzel2019metalearners} to use the linear aggregation function and visualize several trees. This allows us to identify groups of potential voters with similar voting behavior, make predictions of their future voting propensity, and see the effects the different mailers have on future voter turnout.

\begin{figure}[H]
    \centering
    \includegraphics[width=\linewidth]{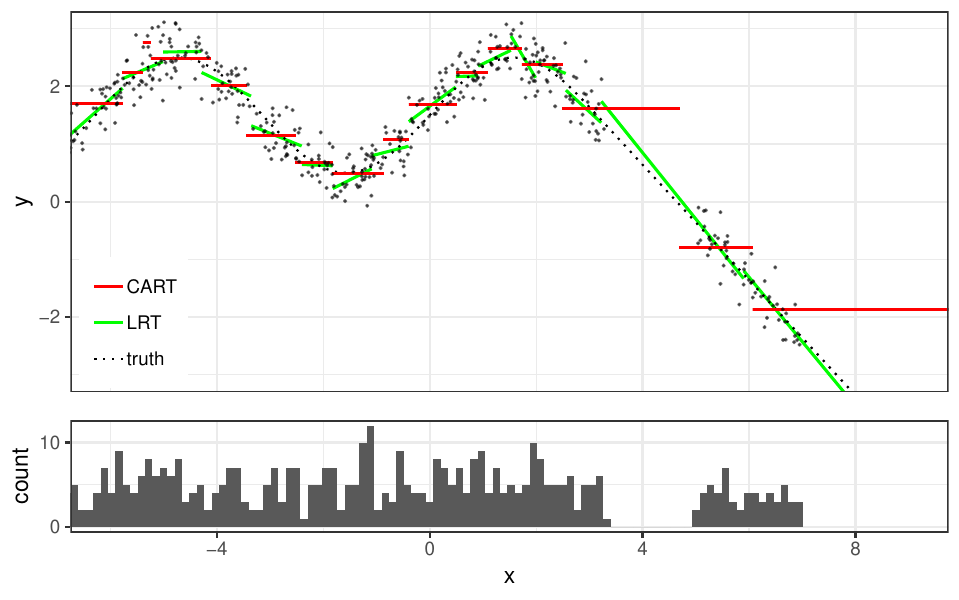}
    \caption{
        Comparison of the classical CART and LRT. The top of the figure shows the fit of Linear Regression Trees and the fit of classical CART, and the lower part shows the density of the training set.  
    }
    \label{fig:classicalVSlinearCART}
\end{figure}

\subsection{Literature}
Studying regression trees with linear aggregation functions has a long history. The earliest reference we are aware of is by \cite{breiman1976general} which precedes the original CART publication \citep{breiman1984classification}. 
The algorithm can be interpreted as a tree algorithm: at each node, the algorithm attempts to introduce a new split along a randomly oriented plane. It then fits a linear model in both children, and it accepts the split if it sufficiently reduces the residual sum of squares as captured by an F-ratio test. If the split is not accepted, the algorithm attempts more splits a limited number of times before declaring the node a leaf node. 

There has been much research on tree-based estimators with linear aggregation functions that builds on this first algorithm, and many different variants of it have been proposed. 
Apart from a few exceptions \citep{zhu2015reinforcement}, the trees are constructed in a recursive strategy and the main differences between the algorithms are, how the splits are generated---the splitting criteria---and when splitting is halted for a node to be defined as a leaf---the stopping criteria. 

For the splitting criteria, there have been many different suggestions. \cite{Torgo:1997} spans the trees similar to the standard CART algorithm without adaptations to the new aggregation function.
\cite{chaudhuri1994piecewise} and \cite{li2000interactive}, on the other hand, fit a linear model, or more generally, a polynomial model, on all units in the current node and then suggest a split by looking at the residuals of the fitted model. 
\cite{zeileis2008model} and \cite{rusch2013gaining}, in turn, use a two-step procedure to find a good split. In the first step, the algorithms fit a linear model on all units in the current node, and they then screen for good splitting covariates using a test statistic that can be computed relatively fast. In a second step, the algorithm then exhaustively searches through all possible splits along the selected features. 

Early stopping and pruning of the trees after they have been trained have both been used for regression trees with linear aggregation functions. 
For example, \cite{quinlan1992learning} and \cite{gama2004functional} build linear models in the pruning phase of each leaf, while \cite{chaudhuri1994piecewise} use a stopping criterion that is based on cross-validation and attempts to estimate whether a further split sufficiently improves the mean squared error. 

In contrast to these existing methods, our stopping criterion accounts for the linear regressions which are fit in the child leaves, and if their combined $R^2$ does not improve on that of the parent node by some amount, the split is rejected.
This both incorporates the linear regressions used in the final aggregation, and allows the depth of the trees to be adaptive to the structure of the data.

While it is always possible to bootstrap these algorithms and combine them with ideas introduced in Random Forests \citep{breiman2001random}, these ideas were mainly studied as regression trees, even though one would expect better predictive power and smoother prediction surfaces in the bagged versions \citep{buhlmann2002analyzing}. Additionally, some of these algorithms would be computationally too expensive to be used in a bagged version.

However, there has been some work done combining RF and bagged trees with linear models. \cite{bloniarz2016supervised} interpret random forests as an adaptive potential nearest neighbor method, following ideas by \cite{hothorn2004bagging}, \cite{lin2006random}, and \cite{meinshausen2006quantile}. Their method, SILO (Supervised Local modeling), uses the random forests algorithm to provide a distance measure, $w$, based on the proximity distance of random forests.
It then defines the random forests prediction, $\hat g(x)$, as a local linear model \citep{stone1977consistent} via the following equations,
\begin{equation} \label{eq:RFLocalModel}
    \begin{aligned}
            \hat f_x(\cdot) &= \argmin_{f \in \mathcal{F}} \sum_{i=1}^N w(x_i, x) (y_i - f(x_i - x))^2,\\
            \hat g(x) &= \hat f_x(0).
    \end{aligned}
\end{equation}
Here, $\mathcal{F}$ is some set of functions. In their paper, they focused in particular on linear models and demonstrated its superior performance over other local models such as LOESS \citep{cleveland1988locally} and untuned RF.

In a recent paper, \cite{friedberg2018local} also combined random forests with linear models. Their work is similar to that of  \cite{bloniarz2016supervised} in the sense that they also use the proximity weights of random forests to fit a local linear model.
Specifically, they focus on linear models, and they fit a ridge regularized linear model, 
\begin{equation}\label{loss}
\begin{pmatrix} 
    \hat{\mu}(x) \\ 
    \hat{\theta}(x) 
\end{pmatrix} 
=
\argmin_{\mu,\theta} \left\{
        \sum_{i=1}^n w(x_i, x)  (y_i - \mu - (x_i - x)^t\theta)^2 + \lambda ||\theta||_2^2
    \right\},
\end{equation}
and use $\hat \mu(x)$ as the estimate for $\E[Y|X = x]$. Similar to \cite{chaudhuri1994piecewise}, they adapt the splitting function to account for the fitting of such local models. When evaluating a split on a parent node P, the potential split is evaluated by fitting a ridge regression in each parent node P to predict $\hat{Y}_{P,Ridge}$ from $X_{P}$ and then using the standard CART split point selection procedure on the residuals $Y_P - \hat{Y}_{P,Ridge}$. 
This results in a fast splitting algorithm which utilizes the residuals to model for local effects in forest creation.
In order to emphasize the difference between this recursive splitting scheme and the scheme presented in this paper, we explore a simple example in Appendix \ref{appendix:LinSplitComparison}. 
By contrast, our splitting algorithm finds the optimal binary split which will result in the minimum training MSE when linear aggregation is used. 
We believe this makes it well suited to settings in which there are both discontinuities and smooth effects in the underlying data.

\subsection{Our Contribution}
Our main contribution is to develop a fast algorithm to find the best partition for a tree with a ridge regularized linear model as aggregation function. 

In Section \ref{section:Algorithm}, we explain the algorithm and we show that its run time is $\bigO(m(n\log(n) + n p^2))$ where $n$ is the number of observations in the current node, $p$ is the number of dimensions that is fit with a linear model, and $m$ is the number of potential splitting covariates. 
In Section \ref{section:simulationstudies}, we use the splitting algorithm with random forests, and we show that it compares favorably on many data sets. Depending on the chosen hyperparameters, the LRF algorithm can mimic either the default RF algorithm or a ridge regularized linear model. Because the algorithms can be trained relatively fast, we were able to connect them with a tuning algorithm, and show how it can quickly adapt to the underlying smoothness levels in different data sets. 
In Section \ref{section:interpretability}, we then apply a single LRT to a real data set to show how its simple structure leads to a deeper understanding of the underlying processes.

\section{The Splitting Algorithm} \label{section:Algorithm}
Random forests are based on regression tree algorithms, which are themselves based on splitting algorithms. These splitting algorithms are typically used to recursively split the space into subspaces which are then encoded in binary regression trees. In this section, we will introduce a new splitting algorithm and discuss its asymptotic runtime.

To describe our novel splitting algorithm, we assume that there are $n$ observations, $(X_i, Y_i)_{i = 1}^{n}$, where $X_i \in \R^d$ is the $d$-dimensional feature vector, and $Y_i \in \R$ is the dependent variable of unit $i$. For a feature $\f \in \{0, \ldots, d\}$, the goal is to find a splitting point $s$ to separate the space into two parts,
\begin{align}
L \define \{x \in \R^d : x[\f] < s\},&  &R \define \{x \in \R^d : x[\f]  \ge s\}.
\end{align}
We call $L$ and $R$ the left and right sides of the partition respectively. 

In many tree-based algorithms, including CART, RF, and GBM, the splitting point $s^*$ is a point that minimizes the combined RSS when fitting a constant in both sides of the potential split,
\begin{equation}\label{eq:meanSplittingProblem}
s^* \in \argmin_{s} \sum_{i : X_i \in L} \Big(Y_i - \Ybar^L\Big)^2 + \sum_{i : X_i \in R} \Big(Y_i - \Ybar^R\Big)^2.
\end{equation}
Here $\Ybar^L$ is the mean of all units in the left partition and $\Ybar^R$ is the mean of all units in the right partition. Note that $L$ and $R$ depend on $s$. 

We generalize this idea to Ridge regression \citep{hoerl1970ridge}. We want to find $s^*$ that minimizes the overall RSS when---in both partitions---a Ridge-regularized linear model is fitted instead of a constant value, 
\begin{equation}\label{eq:ridgeSplittingProblem}
 s^* \in \argmin_s \sum_{i : X_i \in L} \Big(Y_i - \hat Y^L_i\Big)^2 + \sum_{i : X_i \in R} \Big(Y_i - \hat Y^R_i\Big)^2.
\end{equation}
$\hat Y^L_i$ is the fitted value of unit $i$ when a Ridge regression is fit on the left side of the split, and similarly, $\hat Y^R_i$ is the fitted value of unit $i$ when a Ridge regression is fit on the right side of the split.

\subsection{Ridge Splitting Algorithm}
To find an $s^*$ that satisfies (\ref{eq:ridgeSplittingProblem}), it would be enough to exhaustively search through the set $S = \{X_i[\f] : 1 \le i \le n\}$.
However, there are $n$ potential splitting points. Naively fitting a ridge regression to both sides for each potential splitting point requires at least $\Omega(n^2d^2)$ many steps. This is too slow for most data sets, since the splitting algorithm is applied up to $d n$ times for a single regression tree. 
In order to remedy this, we have developed an algorithm for evaluating all of the potential splitting points in a quasilinear runtime in $n$. 

Below we outline the algorithm for finding the optimal splitting point for a continuous feature; the procedure for a categorical feature is outlined in Appendix \ref{appendix:categorical_split}. 

Specifically, the algorithm first sorts the values of feature $\f$ in ascending order. To simplify the notation, let us redefine the ordering of the units in such a way that 
$$
X_1[\f] < X_2[\f] < \cdots < X_n[\f].
$$
Such a sorting can be done in $\O( n \log n)$.\footnote{We assume here that $X_i[\f] \neq X_j[\f]$ for all $j \neq j$. We refer to our implementation in \forestry \citep{forestry} for the general case.}

The algorithm then evaluates the $n-1$ potential splitting points that lie exactly between two observations,
$$
	s_k \define \frac{X_k[\f] + X_{k + 1}[\f]}{2}, \mbox{ for } 1\le k \le n-1,
$$
in such a way that each evaluation only requires $\bigO\left(d^2\right)$ many computations.\footnote{Technically, there are only $\bigO\left(p^2\right)$ steps required where $p \le d$ is the dimension of the ridge regression that is fit in each leaf. For many applications the number of linear features can be a lot smaller than $d$ and this significantly speeds up the computation.}

To motivate the algorithm, define 
\begin{align}
L(k) \define \{x \in \R^d : x[\f] < s_k\},&  &R(k) \define \{x \in \R^d : x[\f]  \ge s_k\}, 
\end{align}
and the RSS when splitting at $s_k$ as $\RSS(k)$, 
\begin{equation}\label{eq:RSS(k))}
\mbox{RSS}(k) \define \sum_{i : X_i \in L(k)} \Big(Y_i - \hat Y^{L(k)}_i\Big)^2 + \sum_{i : X_i \in R(k)} \Big(Y_i - \hat Y^{R(k)}_i\Big)^2.
\end{equation} 
We are then interested in finding:
\begin{equation}\label{eq:minimizing_for_k}
k^* = \argmin_k \mbox{RSS}(k).
\end{equation}
The RSS can now be decomposed into the following terms:
\begin{align*}
\RSS(k)
=  
\Phi_{L(k)} + \Phi_{R(k)} + \sum_{i = 1}^n Y_i ^2.
\end{align*}
Here we use the definition that for a set $H \subset \R^d$,
\begin{align*}
    \Phi_H &\define S_{H}^T A_{H}^{-1} 
        G_{H}^{\phantom{-}}
        A_{H}^{-1} S_{H}^{\phantom{-}}
        - 2 S_{H}^T A_{H}^{-1} S_{H}^{\phantom{-}},\\
     A_H&\define \sum_{i : X_i \in H}
				\begin{bmatrix}X_i \\1 \end{bmatrix}
				\begin{bmatrix}X_i^T&1\end{bmatrix} 
				+
				\lambda 
				\begin{bmatrix} 
				\mathcal{I}_d & 0 \\
				0 & 0
				\end{bmatrix},\\
	S_H &\define \sum_{i : X_i \in H}Y_i \XiO,\\
	G_H &\define \sum_{i : X_i \in H}\XiO \XiOT.
\end{align*}

Our algorithm then exploits the fact that for $H=L_k$ or $H=R_k$, each of these terms can be computed in $\bigO(d^2)$ time based on the terms for $H=L_{k-1}$ and $H=R_{k-1}$.
For example, $S_{L(k)}$ can be computed in $2d$ many computations based on $S_{L(k-1)}$ using the following trivial identity,
$$S_{L(k)} = S_{L(k-1)} + y_k \begin{bmatrix}x_{k} \\1\end{bmatrix}.$$
The  updates for the other terms are presented in Appendix \ref{Appendix_runtime_and_algo}.
The algorithm thus computes $RSS(1), \ldots, RSS(n)$ in sequence from $k=1$ to $n$, and it is thus possible to jointly compute all $RSS(i)$ in $\bigO(nd^2)$ many steps\footnote{We note that this style of algorithm could be adapted to find optimal splits for different parametric aggregation functions. Following the method in \cite{Ghaoui_2009}, one could devise an algorithm using LASSO as the aggregation function, and similar ideas could be used to implement different GLMs for aggregation.}. 
This is a significant improvement over a computationally infeasible naive implementation that would have a runtime of $\Omega(n^2d^2)$ 
\footnote{In order to see the runtime improvement offered by our fast splitting algorithm, we run a timing simulation in Appendix \ref{appendix:runtime} and find that the speedup offered by our fast splitting algorithm is crucial to making Linear Regression Trees feasible for use on even moderately sized data sets.}.

\begin{theorem}
\label{thm:runtime_result}
The runtime for computing $\left\{RSS(1), \ldots, RSS(n)\right\}$ and finding the best splitting point along one variable has a runtime $\bigO\left(n \log n + n d^2\right)$.
\end{theorem}

We prove this Theorem in Appendix \ref{Appendix_runtime_and_algo}. In the appendix, we also elaborate a stopping criteria that we have found to be useful to create smaller trees while improving prediction accuracy and computation speed.

\subsection{Early Stopping Algorithm}
\label{sec:early_stopping}
Although regression trees are prized for their predictive performance, practitioners have long grappled with improving their runtime and interpretability.
It is common for regression trees---especially when part of an ensemble---to be grown to purity (where each leaf node contains only one observation), however this can lead to trees which overfit the training data.
A variety of methods have been developed in order to preserve the predictive strength of regression trees while modifying their structure to be simpler, faster to train, and give more stable predictions.

The post-pruning approach grows a full depth tree, and then iteratively merges child nodes in a way to minimize the empirical error and complexity (number of nodes) of the resulting tree.
When an independent data set is used to prune an existing tree, \cite{nobel_2002_pruning} provides bounds for the performance of pruned classification trees, and \cite{Klusowski_2020_sparsecart} provides bounds for the performance of pruned regression trees. \cite{Klusowski_2020_sparsecart} also show that for sparse additive outcomes, a CART tree pruned in this way will asymptotically recover the true sparse structure.

An alternative approach limits the depth of trees during construction rather than after a full tree has been constructed.
The Bayesian Additive Regression Trees \citep{chipman2007,chipman2010bart} algorithm limits the depth of individual regression trees by placing an exponential prior on the depth of each tree, limiting the depth drastically, but allowing deeper trees when such a tree fits the data well.
The RuleFit algorithm \citep{friedman_2008} creates an ensemble of trees where each tree is given a different random maximum depth, and all splits that would make the tree deeper than this depth are automatically rejected.

We introduce the early stopping algorithm in order to provide simpler trees, a runtime improvement on large data sets, and larger leaf nodes---without reducing the adaptivity or predictive power of the trees.
Unlike pruning procedures, our algorithm rejects potential splits during runtime, which leads to a large speedup over creating full trees.
Our algorithm also varies from those which limit depth, node sizes, or node counts across the tree as it allows for some branches of the tree to grow quite deep, and other branches to grow much shorter, based on the improvement in predictive accuracy brought by further splits.

The early stopping algorithm follows the following steps (for details, see Algorithm \ref{algo:EarlyStopping}):
\begin{enumerate}
    \item Once the optimal split for a parent node is calculated, the current observations for the node are split into $k$ folds.
    \item For each fold, the predictions---using a regularized linear model---are made for both the parent, and the two child nodes which result from the split. 
    \item The $R^2$ of the parent node regression is compared to the $R^2$ of the two child nodes, if the two resulting linear models increase the $R^2$ for the current node's observations by a pre-specified percentage, the potential split is accepted, otherwise, the split is rejected, and the parent node becomes a leaf node.
\end{enumerate}
As the overall runtime of creating a decision tree of depth $D$, which samples \textit{mtry} potential splitting features at each node is $\Theta \left(D * \textit{mtry}* \left(n \log{n} + n d^2 \right)  \right)$, decreasing the depth of potential trees (or even several branches of the tree) has a large effect on the overall runtime of the tree.
This provides a runtime advantage for early stopping which can be very impactful on large data sets.
As the early stopping algorithm prohibits further splits which do not increase the $R^2$ of the fit, this can also help the tree avoid continued splitting when there are no informative splits available.
This is important as it allows the leaf node sizes to remain large while removing uninformative splits which might have caused the tree to overfit the data.

\section{Predictive Performance} \label{section:simulationstudies}
In this section, we will evaluate different tree-based estimators with linear aggregation functions and some natural baseline estimators. 
It is well known that RF outperforms regression trees, and we have therefore implemented a version of RF that utilizes our new splitting algorithm. We call this Linear Random Forests (\LRF) and we demonstrate its predictive performance in the following. 
A version of LRF is implemented in our \forestry package \citep{forestry}. For further details of the software package, see Appendix \ref{appendix:software}

\subsection{Methods and Tuning}
We compare the predictive performance of LRF with seven competitors:
\begin{itemize}
    \item The random forest algorithm as implemented in the \ranger package \citep{wright2015ranger} and in the \forestry package.
    \item The Bayesian Additive Regression Trees (BART) \citep{chipman2007,chipman2010bart} algorithm as implemented in the \dbarts package \citep{dorie2020dbarts}.
    \item Local linear forests \citep{friedberg2018local} as implemented in the \grf package \citep{athey2019generalized}.
    \item {The Rule and Instance based Regression Model presented in \cite{quinlan1992learning} as implemented in the \Cubist package.}
    \item Generalized Linear Models as implemented in the \glmnet package \citep{friedman2010regularization}. 
    \item The RuleFit algorithm \citep{friedman_2008} as implemented in the \textbf{pre} package \citep{Fokkema_2020}. 
    \item Gradient Boosting \citep{friedman2001greedy} as implemented in the \textbf{gbm} package \citep{ridgeway_2004}.
\end{itemize}

In most real world prediction tasks, appliers carefully tune their methods to get the best possible performance. We believe that this is also important when comparing different methods. That is why we use the \caret package \citep{kuhn2008building} to tune the hyperparameters of all of the above methods.
\footnote{BART often achieves excellent performance without tuning \citep{hill2011bayesian,dorie2018automated, hill2020bart}. But we also tune BART because its performance improves with tuning in these simulations.}
Specifically, we use the \caret package's adaptive random tuning over 100 hyperparameter settings, using 8-fold cross-validation tuning repeated and averaged over four iterations.
The hyperparameters tuned for each algorithm are presented in Table \ref{tab:hyperparameters_tuned} in Appendix \ref{appendix:hyperparams}. 
As with the other hyperparameters, we use \textbf{caret} to tune the value of $\lambda$ used in LRF. 
For more details, see Appendix \ref{appendix:optimal_lambda}.

\subsection{Experiments}
LRF is particularly well suited for picking up smooth signals, however, depending on the hyperparameter choice, it can behave similar to a regularized linear model, a classical RF, or a mixture of the two.
To demonstrate this adaptivity, we first analyze artificially generated examples. We will see that LRF automatically adapts to the smoothness of the underlying data generating distribution. 
In the second part of this study, we show the predictive performance of our estimator on real world data.

\subsubsection{Adaptivity for the Appropriate Level of Smoothness}

We analyze three different setups. In the first setup, we use linear models to generate the data. Here we expect \glmnet to perform well and the default versions of RF as implemented in \ranger and \forestry to not perform well. 
In the second setup we use step functions. Naturally, we expect RF and other tree-based algorithms to perform well and \glmnet to perform relatively worse. In the third setup, there are areas that are highly non-linear and other areas that are linear. It will thus be difficult for both the default RF algorithms and \glmnet to perform well on this data set, but we expect the adaptivity of LRF to excel here. 

\begin{enumerate}
	\item \textbf{Linear Response Surface} \\
	In \textbf{Experiment 1}, we start with a purely linear response surface. 
	We simulate the features from a normal distribution and we generate $Y$ according to the following linear model:
	\begin{equation*}
	\begin{aligned}
	&X \sim N(0,1) \in \R^{10},~\varepsilon \sim N(0,4)\\
	&Y = f_L(X) + \varepsilon,\\
	&f_L(X) = -0.47 X_2 -0.98 X_3 - 0.87 X_4 + 0.63 X_8 - 0.64 X_{10}.
	\end{aligned}
	\end{equation*}
	
	\item \textbf{Step Functions}\label{sec:simstep}\\
	Next, we are interested in the other extreme. 
	In \textbf{Experiment 2}, we simulate the features from a 10-dimensional normal distribution, $X \sim N(0,1) \in \R^{10}$, and we create a regression function from a regression tree with random splits, 50 leaf nodes, and randomly sampled response values between -10 and 10 and we call this function $f_S$.
	$Y$ is then generated according to the following model:
	\begin{equation*}
    	\begin{aligned}
    	&Y = f_S(X) + \varepsilon\\
    	&\varepsilon \sim N(0,1)
    	\end{aligned}
	\end{equation*}
	A specific description of how $f_S$ was generated can be found in detail in Appendix \ref{sec:AppCreatestepResSurf}.
	
	\item \textbf{Linear Function with Steps}\\
	Finally, we look at a data set that exhibits both a linear part and a step-wise constant part. Specifically, we split the space into two parts and we combine the regression functions from \textbf{Experiment 1} and \textbf{Experiment 2} to create a new regression function as follows:
	\begin{equation*}
    	\begin{aligned}
        	&Y = \begin{cases} f_L(X)&\mbox{if } X_1 < .5\\
        	                  f_S(X) &\mbox{else}
        	     \end{cases} +\varepsilon\\
        	&\varepsilon \sim N(0,1).
    	\end{aligned}
	\end{equation*}
\end{enumerate}	

\begin{figure}
	\begin{center}
	\begin{subfigure}{.495\textwidth}
		\centering
		\includegraphics[width=\linewidth]{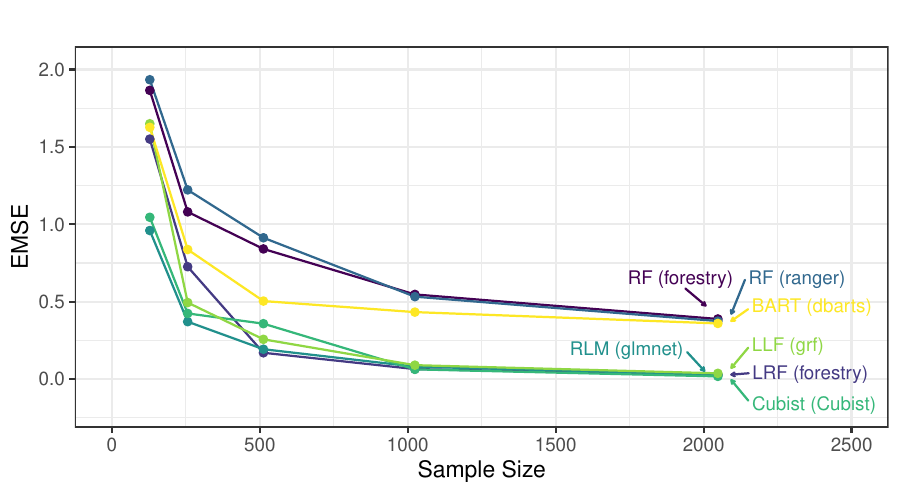}
	\end{subfigure} 
	\begin{subfigure}{.495\textwidth}
		\centering
		\includegraphics[width=\linewidth]{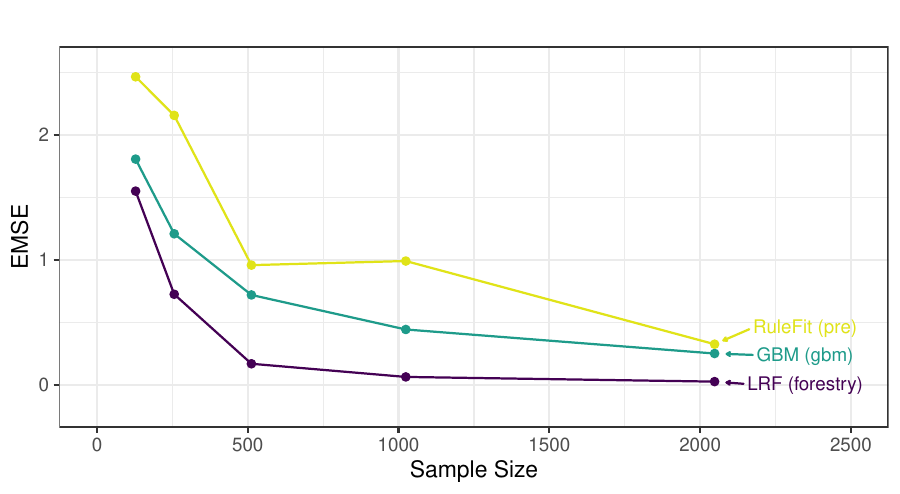}
		\vspace*{-6mm}
	\end{subfigure} 
	\\(a) Experiment 1: Linear regression function.
    \label{fig:Exp1}
	\end{center}

    \begin{center}
	\begin{subfigure}{.495\textwidth}
		\centering
		\includegraphics[width=\linewidth]{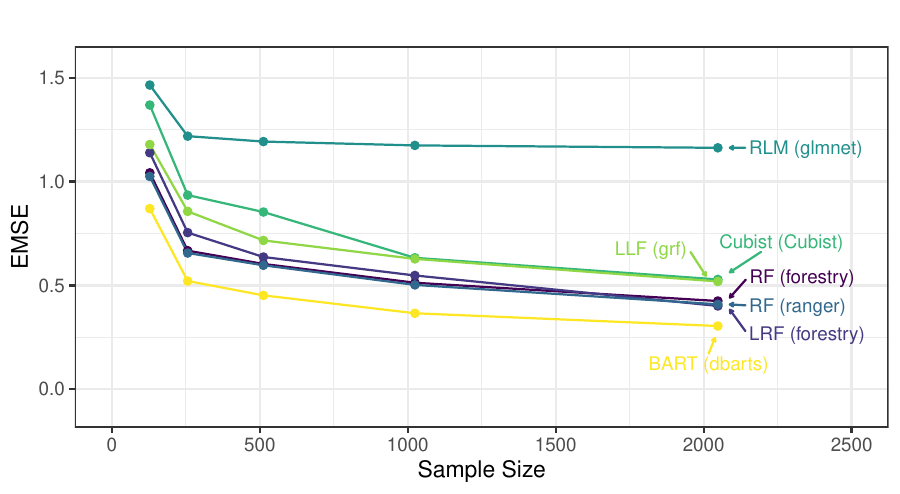}
	\end{subfigure} 
		\begin{subfigure}{.495\textwidth}
		\centering
		\includegraphics[width=\linewidth]{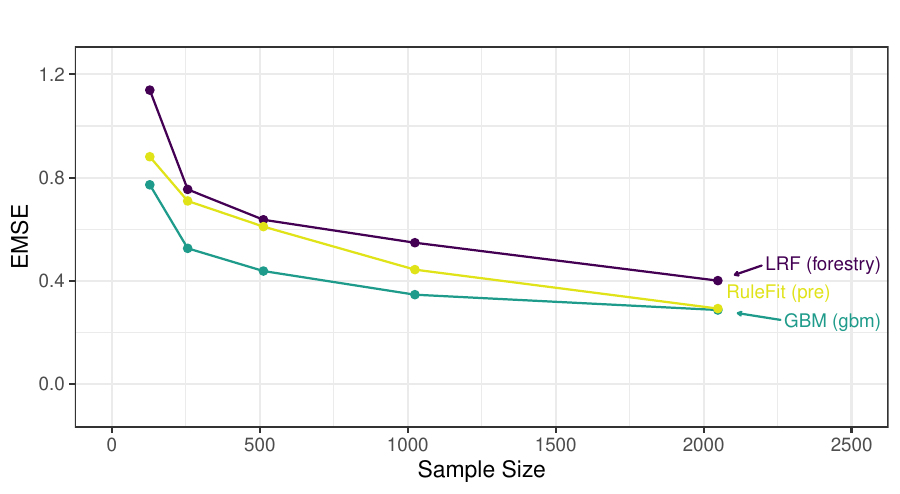}
		\vspace*{-6mm}
	\end{subfigure} 
	\label{fig:Exp2}
	\\ (b) Experiment 2: Step function.
	\end{center}
	
	\begin{center}
	\begin{subfigure}{.495\textwidth}
		\centering
		\includegraphics[width=\linewidth]{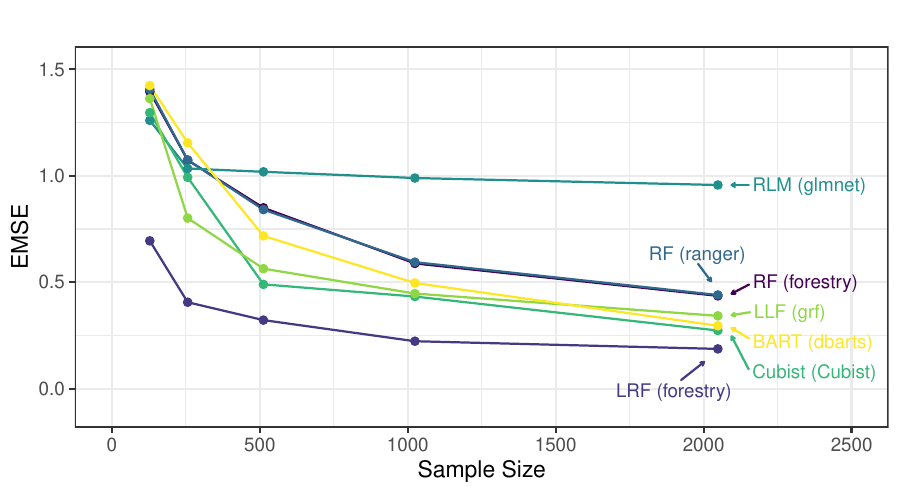}
	\end{subfigure}
	\begin{subfigure}{.495\textwidth}
	\centering
	\includegraphics[width=\linewidth]{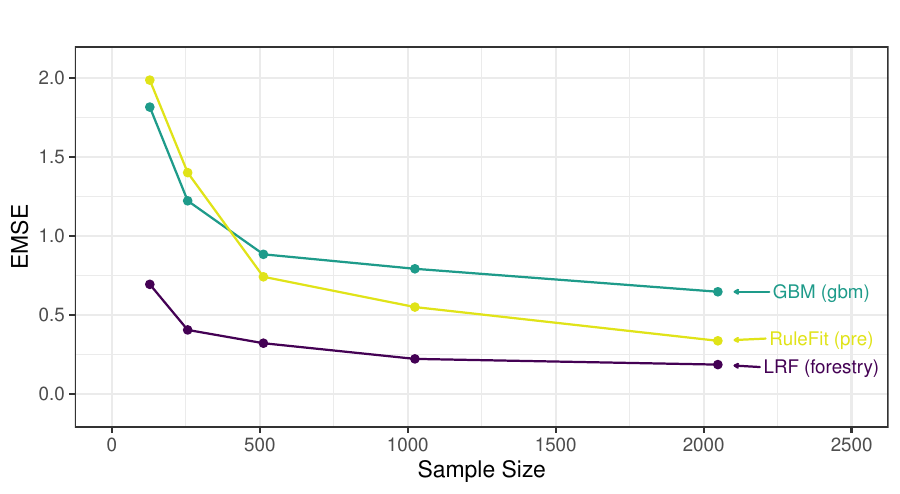}
	\vspace*{-6mm}
	\end{subfigure} 
	\\ (c) Experiment 3: Partly linear and partly a step function.
	\label{fig:sfig2}
\end{center}
	\caption{Different levels of smoothness. In Experiment 1, the response surface is a linear function, in Experiment 2, it is a step function, and in Experiment 3, it is partly a step function and partly a linear function.}
	\label{fig:MSE_simulation}
\end{figure}

The results of these simulations for a range of sample sizes can be found in Figure \ref{fig:MSE_simulation}.
For each training sample size, the EMSE was calculated over a 10,000 point test set drawn from the same data generating process.

\subsubsection{Smoothness Simulation Results}
\label{smoothness_simulation_results}
In Experiment 1, RF and BART perform relatively poorly estimating a purely linear response. 
In contrast, the algorithms designed to handle smoothness do quite well here, with Cubist, Local Linear Forests, Linear RF, and regularized linear models all performing quite well.

In the second experiment, where we have a highly non-linear response, BART performs very well, followed by RF and Linear RF. 
Although the performance of Linear RF closely trails that of RF, performance of the other algorithms exploiting smoothness fall behind here. 

Experiment 3 contains a response function which is in some local neighborhoods linear and in others a step function. We see that Linear RF does quite well compared to both the tree based methods such as RF and BART, and the models which utilize smoothness.

Through examining three simple experiments with varying degrees of smoothness, we see that Linear RF can be quite adaptive to the underlying smoothness in the response function. This allows Linear RF to act as a Random Forest, a regularized linear model, or anything in between.

\subsubsection{Real World Data Sets}
\label{sec:realworlddatasets}
To analyze the behavior of these estimators on real data, we used a variety of data sets.
Table \ref{tbl:dssummary} describes the metrics of the data sets that we have used and Table \ref{fig:realperformance} shows the performance of the different estimators. 

Of these data sets, \emph{Autos} describes pricing data and many related features scraped from German Ebay listings, \emph{Bike} describes Capital Bikeshare user counts and corresponding weather features.
These data sets are available on Kaggle \citep{usedcars} and the UCI repository respectively \citep{bike}.
The remaining data sets were taken from Breiman's original regression performance section \citep{breiman2001random}, with the test setups altered only slightly. 

In order to evaluate the performance of the different models, we used a hold out data set as test set for the \emph{Abalone}, \emph{Autos}, and \emph{Bike} data set.
The \emph{Boston}, \emph{Ozone}, and \emph{Servo} data sets are relatively small and we thus used 5-fold cross validation to estimate the Root Mean Squared Error (RMSE). For the three Friedman simulations, we kept Breiman's original scheme by using 2000 simulated test points.

We include a variety of data sets which have a range of dimensions and noise levels to compare estimators across a range of scenarios practitioners might encounter in real data problems.
It is well known that bagging estimators perform variance reduction on the base learners they utilize. 
With this in mind, we recommend bagging Linear Regression Trees in data examples where the noise is high.

\subsection{Results}

\begin{table}[ht]
\begin{center}
\begin{tabular}{rrrrr}
  \hline
name & ntrain & ntest & dim & numeric features \\ 
  \hline
Abalone & 2089 & 2088 & 11 & 11 \\ 
  Autos & 1206 & 39001 & 12 & 6 \\ 
  Bike & 869 & 16510 & 16 & 16 \\ 
  Boston & 506 &  & 9 & 9 \\ 
  Friedman 1 & 1000 & 2000 & 11 & 11 \\ 
  Friedman 2 & 1000 & 2000 & 5 & 5 \\ 
  Friedman 3 & 1000 & 2000 & 5 & 5 \\ 
  Ozone & 330 &  & 9 & 9 \\ 
  Servo & 167 &  & 13 & 13 \\ 
   \hline
\end{tabular}
\caption{Summary of real world data sets.} 
\label{tbl:dssummary}
\end{center}
\end{table}

\begin{table}[H]
    \centering
    \includegraphics[width =1\textwidth]{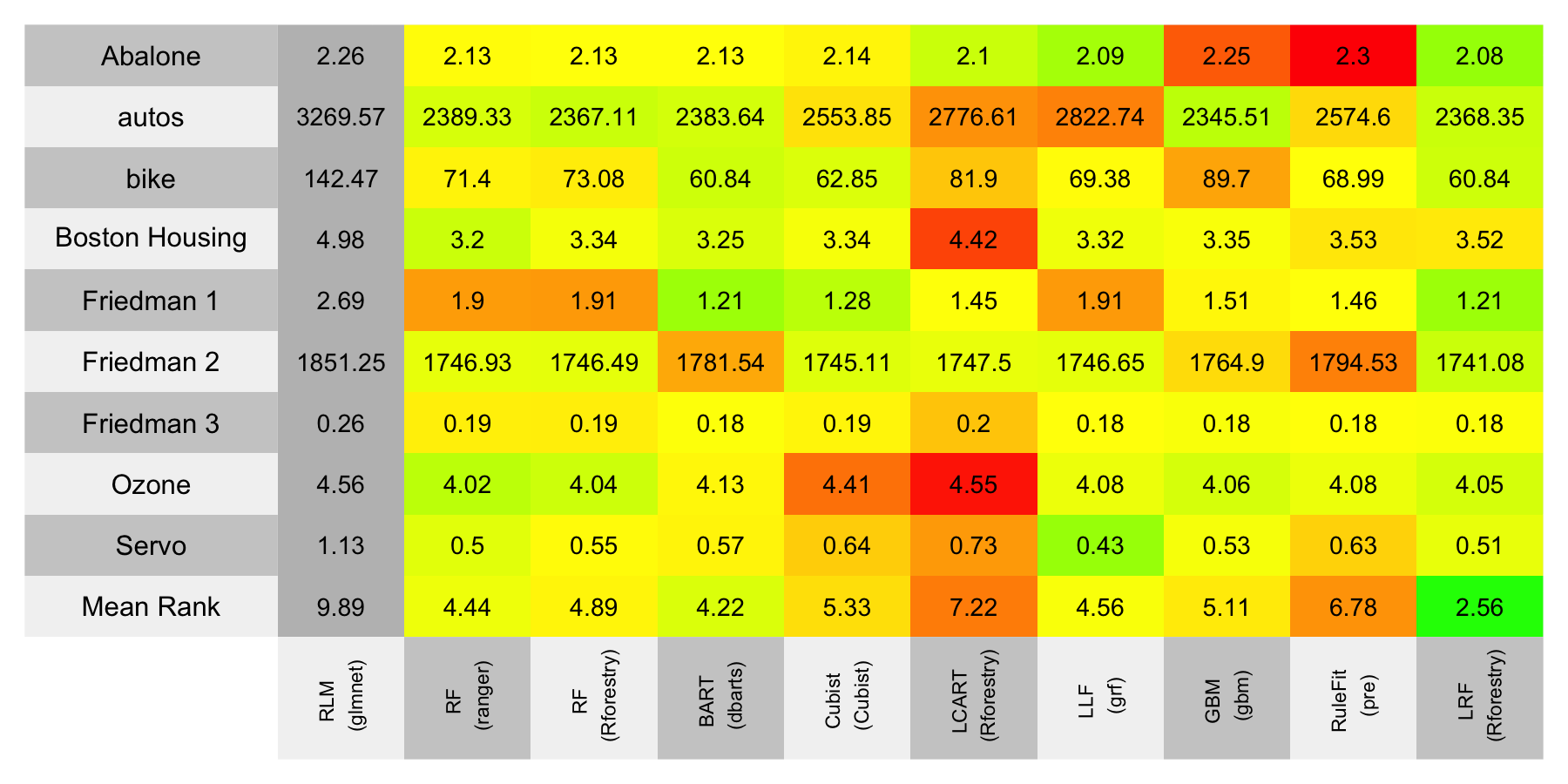}
    \caption{\textit{Estimator RMSE compared across real world data sets. The final row displays the mean performance rank across the different data sets for each estimator.}}
    \label{fig:realperformance}
\end{table}

Figure \ref{fig:MSE_simulation} shows the results for the different estimators on the simulated response surfaces detailed in Section \ref{section:simulationstudies}.
As expected, we find that Regularized Linear Models (RLM) as implemented in the \glmnet package perform very well on Experiment 1 where the response is linear, while it performs poorly on Experiments 2 and 3. 
The default random forest estimators implemented for example in \ranger performs very well on Experiment 2, but it performs rather poorly on Experiment 1 and 3, as estimators which can exploit the linearity benefit greatly here. 
The estimators which can pick up linear signals (RLM, LLF, and LRF), perform nearly identically on Experiment 1, showing a tuned LRF can mimic the linear performance of a linear model such as RLM.
Experiment 2 showcases a step function, in which smooth estimators such as RLM and LLF suffer, while RF and boosting algorithms such as BART excel. In this scenario, the performance of LRF now converges to match that of the purely random forest algorithms.
Finally, in Experiment 3, the response surface is evenly distributed between smooth response values, and a step function. In this case, there are areas of weakness for both the smooth estimators and the RF estimators, but the adaptivity of LRF allows it to adapt to the differences across the response surface, and fit the appropriately different estimators on each portion.

The results shown in Table \ref{fig:realperformance} are mixed, but display the adaptivity of LRF on data sets that are linear as well as data sets that are highly nonlinear. 
This adaptivity results in an estimator that can exploit the presence of smooth signals very well and remain competitive over a wide range of response surfaces.

\subsection{High Dimensional Simulation}
\label{sec:high_dim_simulations}
One benefit of using regularized linear models as the aggregation functions is the ability of regularized linear models to shrink the effects of spurious predictors towards zero. 
In this section, we explore a simple simulation where the outcome is a nonlinear function of four covariates and there are many noisy covariates included in the model.

\subsubsection{Data Generating Process}
We follow an example from \cite{athey2019generalized} with the outcome modelled as follows.
First, $X \sim N(0,1)^{100}$, then $Y$ is generated as follows:
\begin{align}
\label{outcome_sim}
    Y = \varsigma\left(X_{1}\right) \varsigma\left(X_{2}\right) + \varsigma\left(X_{3}\right) \varsigma\left(X_{4}\right) + \epsilon
\end{align}
Where:
\begin{align}
    \varsigma(x)=\frac{2}{1+e^{-12(x-.5)}}
\end{align}
And $\epsilon$ is selected to give a signal to noise ratio of 2.
As the true outcome depends on only the first four covariates, the majority of the covariates are noisy. 

\begin{figure}[H]
    \centering
    \includegraphics[width =.7\textwidth]{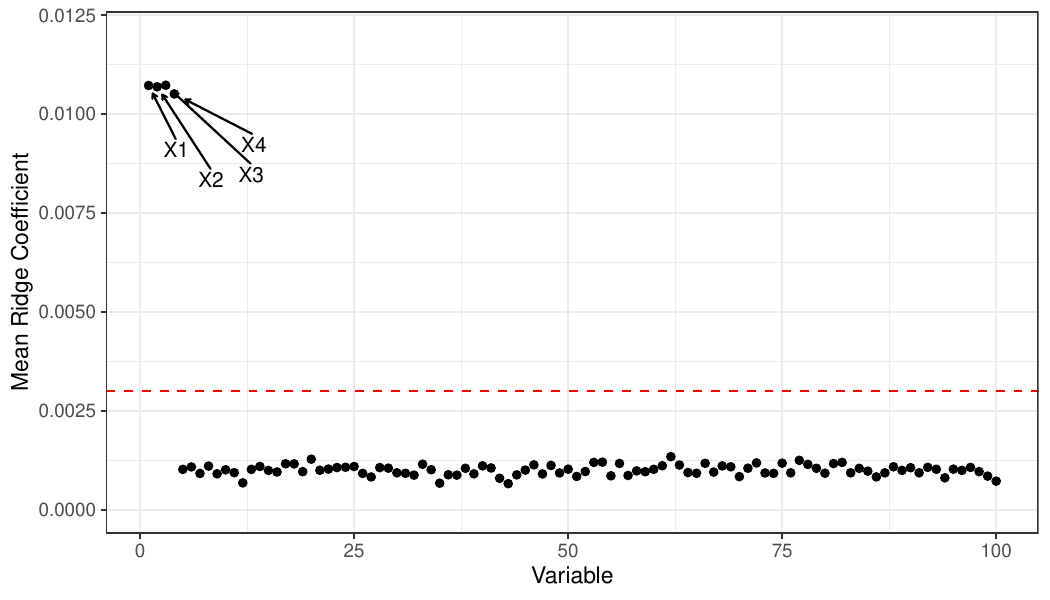}
    \caption{The mean Ridge Coefficients from simulated data generated according to Equation \ref{outcome_sim}, repeated over 100 Monte Carlo replications. The true outcome relies on only the first four covariates, and these coefficients have nonzero slope which is picked up by the Linear Random Forest. The horizontal line corresponds to 1.96 times the sample standard deviation of the coefficients over the Monte Carlo replications.}
    \label{fig:high_dim_simulation}
\end{figure}

\section{Interpretability} \label{section:interpretability}
In this section, we wish to show how linear aggregation can be used to create more interpretable estimators, and we illustrate this by analyzing a large voter turnout experiment. We will show the usefulness of linear aggregation to better understand the heterogeneity in the data. We first introduce the data set used, outline the problem of estimating the Conditional Average Treatment Effect (CATE), and then apply LRF to the data set to estimate the CATE.

\subsection{Social Pressure and Voter Turnout Data Set}
We use a data set from a large field experiment that has been conducted by \cite{GerberGreenLarimer} to measure the effectiveness of four different mailers on voter turnout.
The data set contains 344,084 potential voters in the August 2006 Michigan statewide election with a wide range of offices and proposals on the ballot.

There are four mailers used, each of which is designed to apply a different amount of social pressure to the potential voter in order to incentivize voting in the upcoming election. 
In increasing order of social pressure, the mailers are summarized as follows:
\begin{itemize}
    \item Civic Duty (CD): The civic duty mailer reminds the recipient that it is their civic duty to vote.
    \item Hawthorne (HT): The Hawthorne mailer informs the potential voter that they are being studied in a survey of voter turnout, but reassures them that their decision to vote will remain anonymous.
    \item Self (SE): The self mailer informs the potential voter that the other members of their household will be informed on whether or not they vote in the coming election.
    \item Neighbors (NE): The neighbors mailer informs the potential voters that their neighbors will be informed on whether or not they vote in the coming election.
\end{itemize}

For the full text of each mailer, see \cite{GerberGreenLarimer}.
The data set also contains seven key individual-level covariates: gender, age, and indicators of whether or not the individual voted in the primary elections in 2000, 2002, 2004, or the general election in 2000 and 2002. 
We also derived three further covariates from the ones above:
\begin{itemize}
    \item Cumulative Voting History (CVH): The number of times the voter has voted within the last five elections before the 2004 general election.
    \item Cumulative Primary Voting History (CPVH): The number of primary elections a voter has participated in between 2000 and 2004.
    \item Cumulative General election Voting History (CGVH): The number of general elections a voter has participated in between 2000 and 2002. 
\end{itemize}

\cite{kunzel2019metalearners} analyze the effect of the Neighbors mailer while paying specific attention to uncovering the heterogeneity in the data. 
Specifically, they estimate the Conditional Average Treatment Effect (CATE),
$$
\tau(x) = \E[Y(1) - Y(0)| X = x].
$$
Where, $Y(1) \in \{0,1\}$ is the indicator of voting when a unit is treated (is assigned to receive a Neighbors mailer), and $Y(0)$ is the indicator of voting when the unit is in the control group, and $x$ is the vector of covariates.

Estimating the CATE is useful for targeting treatment to individuals with a particularly strong effect.
\cite{kunzel2018causaltoolbox, kunzel2019metalearners}, for example, uncover the heterogeneity by using partial dependence plots \citep{friedman2001greedy} in combination with looking at specific subgroups that have been defined through subject knowledge and exhaustive exploratory data analysis. This process involved a lot of hand selection of subgroups and covariates of interest and prior knowledge of the causal process. This is a common approach used by substantive researchers to find heterogeneous treatment effects \citep[e.g.,][]{arceneaux2009mobilized,enos2014increasing}.
In contrast to this approach, using a linear aggregation function in RF directly enables us discover subgroups of interest in a data-driven and algorithmic way that is still interpretable.

\subsection{Interpretable Estimators with LRF} \label{sec:SL_interpretability}
Introduced in \cite{kunzel2019metalearners}, the S-Learner estimates the outcome by using all of the features and the treatment indicator, without giving the treatment indicator a special role,
$$
    \hat \mu(x, w) = \hat \E[Y | X = x, W = w].
$$
The predicted CATE for a unit is then the difference between the predicted values when the treatment-assignment indicator is changed from control to treatment, with all other features held fixed, 
\begin{equation} \label{eq:S-Learner_default_first_step}
    \hat \tau(x) = \hat \mu(x, 1) - \hat \mu(x, 0).
\end{equation}
Any machine learning algorithm can be used for the regression step, and here we use LRF.
We additionally make several changes to the standard formulation of the S-Learner outlined in \cite{kunzel2019metalearners}:
\begin{enumerate}
    \item
    As we have four different treatments, we encode a treatment indicator for each of the four treatments and we estimate the response function using all covariates and the treatment indicators as independent variables,
    $$
    \hat \mu(x, w_{1}, \ldots, w_{4}) = \hat \E[Y | X = x, W_1 = w_1, \ldots, W_4 = w_4].
    $$
    The CATE for the CD mailer is then defined as:
    $$
    \hat \tau(x) = \hat \mu(x, 1, 0, 0 ,0) - \hat \mu(x, 0, 0, 0, 0)
    $$
    and the treatment effects for the other mailers are defined equivalently. 
    \item 
    We choose to linearly adapt the four treatment indicators and we allow splits on all covariates except the variables that encode the treatment assignment. This allows splits to uncover heterogeneity in the CATE function, and the linear model in each leaf to contain coefficients corresponding to the treatment effect of each mailer in the current leaf.
    Thus $\hat b$, the intercept of the linear model (\code{untr BL} in the figure), is an estimate for the baseline turnout of voters falling into this leaf and the $\hat \tau_i$ (\code{TE(i)} for $i \in \{CD,HT,SE,NE\}$ in the figure) is an estimate for the treatment effect of the different mailers\footnote{For details, see Appendix \ref{appendix:interpretability}}.
    \item
    We implement the trees with \textit{honesty}. Specifically, we split the training set into two disjoint sets: A splitting set that is used to select splits for the tree structure and an averaging set that is used to compute the linear models in each terminal node.\footnote{This property has been studied before under different names: e.g., \cite{scornet2015consistency, wager2015estimation}.}
    This means the splits and thus the structure of the trees are based on a set that is independent from the set that is used to estimate the linear model coefficients. 
    \item 
    In order to avoid overfitting and create interpretable estimators with performance that is comparable to that of much deeper CART trees, we use the one step look-ahead stopping criteria introduced in Section \ref{sec:early_stopping}.
    This also allows us to fit a linear model on a large number of observations in each leaf, resulting in increased stability of the resulting coefficients.
\end{enumerate}
In practice, the overfit penalty can be chosen by cross validation or taken heuristically to penalize linearity based on subjective knowledge.
Here we take essentially linear splits ($\lambda = 10^{-8}$) as it allows the coefficients of the linear models in each leaf to be interpreted as the average treatment effect of each mailer in the particular leaf \citep{lin2013agnostic}.

By using LRF, this formulation of the S-Learner can take advantage of the linear splitting algorithm in order to evaluate splits on all covariates that result in treatment effect heterogeneity, and LRF can grow much shallower trees which are still optimized for predictive accuracy.
We find in this data set that the Linear Regression Trees displayed below have comparable performance to a standard CART tree that is up to 100 levels deep.\footnote{To reproduce this example, see the replication archive at \url{https://github.com/forestry-labs/RidgePaperReplication}}
Further, we find that the LRF ensemble has base learner predictions which are more highly correlated than those of a standard random forest, implying that the comparative performance of the ensemble relative to the base learner will lower for LRF.

\begin{figure}[ht]
	\begin{center}
		\begin{subfigure}[b]{.895\textwidth}
			\centering
			\includegraphics[height=.78\linewidth]{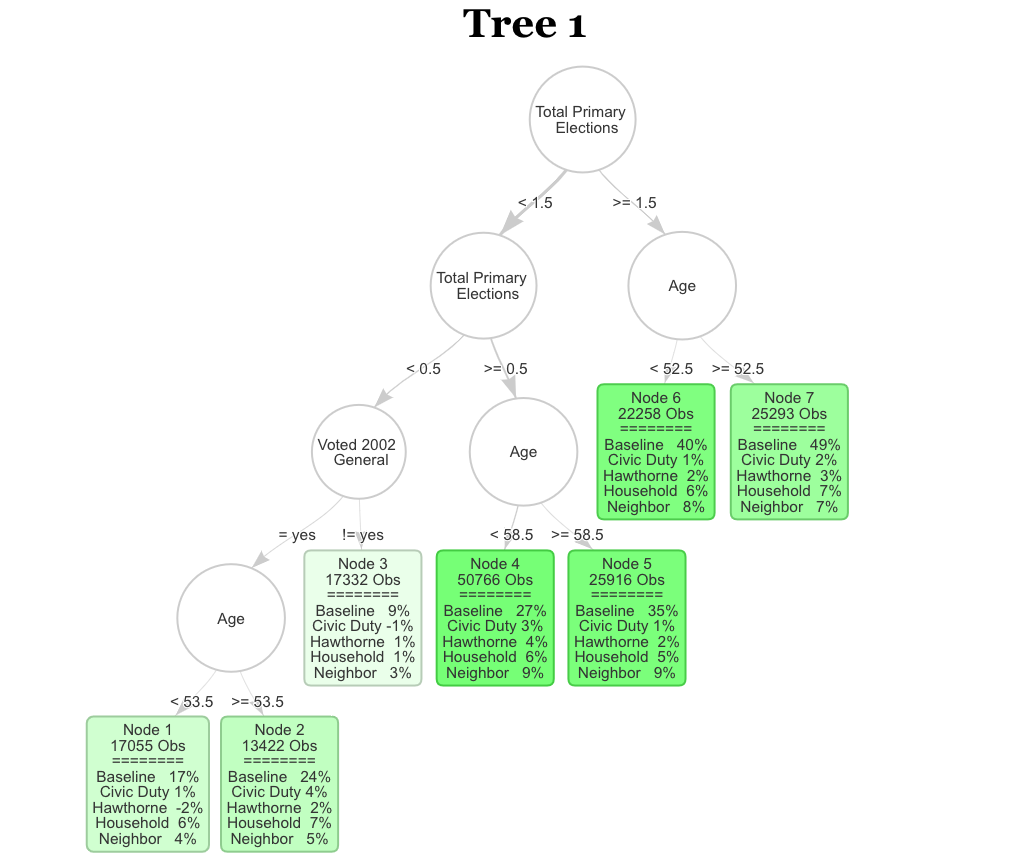}
	    \caption{}
		\end{subfigure} 
	\end{center}
	\label{fig:PlottedTrees}
	\caption{The first tree of the S-Learner as described in Section \ref{sec:SL_interpretability}. The first row in each leaf contains the number of observations in the averaging set that fall into the leaf. The second part of each leaf displays the regression coefficients. 
	\code{Baseline} stands for untreated base turnout of that leaf and it can be interpreted as the proportion of units that fall into that leaf who voted in the 2004 general election. 
    Each coefficient corresponds to the ATE of the specific mailer in the leaf.
	The color strength is chosen proportional to the treatment effect for the neighbors treatment.
	}
\end{figure}

\subsection{Results}
\label{sec_interpretation}

As the LRF trees use linear aggregation functions, the predictions of the ensemble tend to be more correlated than those of a standard random forest.
For this reason, a single linear tree has quite strong performance compared to that of the full ensemble, and can provide a much more interpretable estimator.
In order to understand the form of the CATE function, we study a single tree.
The conclusions still hold based on examining additional trees in the ensemble (shown in Appendix \ref{appendix:interpretability}). 


Examining the results, we find several subgroups for which the various treatments display stark heterogeneity.
There seems to be a subgroup of voters, the voters falling into Node 3, that have a very low previous election participation.
This group has very low baseline turnout, and low treatment effects for all of the mailers.
This could correspond to a disaffected group of voters which have very little interest in participating in the elections under any circumstances.
The low baseline turnout and low treatment effects invite further experimentation on methods of increasing the turnout of this subgroup.

We also find that there is a subgroup of citizens that have voted in at least two out of the three recorded primary elections (Those falling into Nodes 6 and 7). 
These units have a high voting propensity even if they do not receive any of the mailers. The treatment effects are large for this group of high propensity voters. In short, these treatments widen disparities in participation by mobilizing high-propensity citizens more than low-propensity citizens. This may be surprising because one may not expect a letter in the mail to significantly impact the turnout decisions of voters with a high  propensity to vote. 

It turns out that substantive researchers have reached the same conclusion previously using heuristic methods that rely on substantive knowledge for selecting subgroups \citep[e.g.,][]{enos2014increasing}. We were able to find the results using an algorithmic approach. 

Another consistent trend is a number of splits on age around 50-55 years old.
Older voters tend to have a much higher baseline turnout of approximately an additional $10\%$.
This is also in line with known results on voter turnout increasing among older voters.

The simple and expressive form of the tree enables us to find subgroups based on our covariates which display heterogeneity in the treatment effects of the various mailers.
Further, the data driven construction of the tree provides both a strong predictive model, and does not require expert input to specify subgroups of interest, and it uncovers them in an unsupervised way.
This is a strong tool for scientists to use in order to understand large and complex data sets through interpretable and expressive nonparametric models.

\section{Conclusion} \label{section:conclusion}
In this paper, we have proposed three main changes to tree-based estimators and explored their consequences in several experiments, real world data examples, and a runtime analysis. Specifically, our changes were focused on introducing linear aggregation functions for prediction in the leaf nodes of regression trees, modifying the splitting algorithm to efficiently find optimal splits for these aggregation functions, and adding a stopping criteria which limits further node creation when splitting would not increase the $R^2$ by a predetermined percentage.

In 12 different prediction tasks, we found that these new estimators increased the predictive power and that they can adapt to the underlying local smoothness conditions of the data sets (Section \ref{section:simulationstudies}). 
We then illustrated in a real world experiment how these new algorithms lead to more interpretable estimators, and we demonstrated how a visualization of regression trees can be used to perform inference through both the selected splits and resulting regression coefficients in the leaves (Section \ref{section:interpretability}).

We have also implemented several tree-based estimators including, CART (as LCART), random forests (as LRF), and gradient boosting in the \forestry package. In future work, we are interested in extending these ideas to other estimators such as BART. 
We believe that LRF can be less biased than the usual versions of RF, and we wonder whether it would improve nonparametric confidence interval estimation.
Finally, we believe that the regression coefficients could be substantively informative, and we would like to study statistical tests that can be used to determine whether the estimated coefficients for the leaves vary significantly across leaves.

\bibliographystyle{apalike}
\bibliography{references}

\newpage
\appendix

\section{Linear Splits Comparison}
\label{appendix:LinSplitComparison}
In order to illustrate the difference between the recursive splitting function of Local Linear Forests \citep{friedberg2018local} as implemented in the \grf package \citep{athey2019generalized}, and that of Linear Random Forests, as implemented in the \forestry package, we create a simple simulated example to visualize the different splits.

\subsubsection*{Comparing the Selected Split}
\begin{figure}[H]
    \centering
    \includegraphics[width=.8\linewidth]{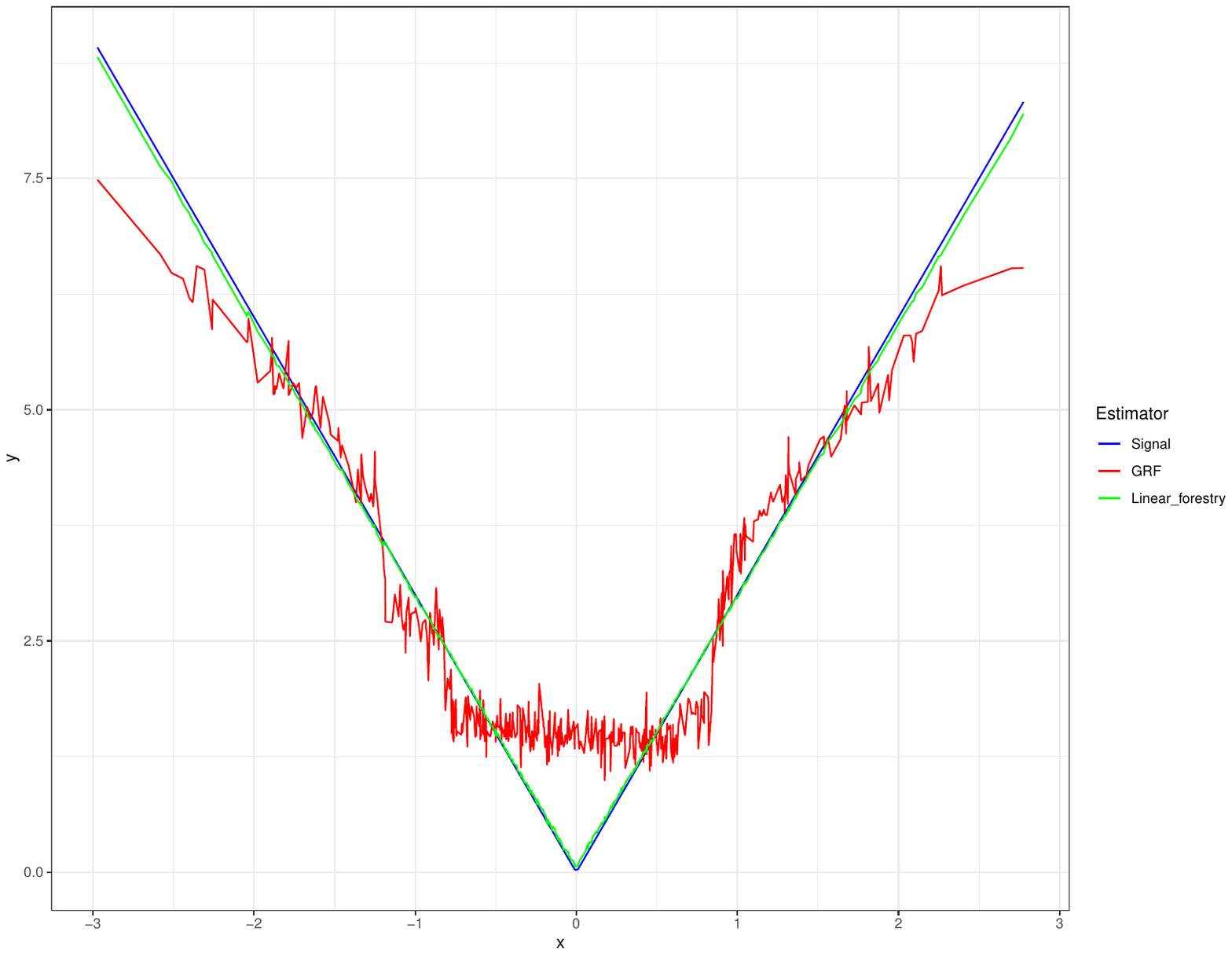}
    \caption{
        This example has a piecewise linear response, with alternating slopes in the first covariate.
        Here we plot $X_1$ against the outcome $y$, and overlay the fitted values which are returned by both the Local Linear Forest and Linear Regression Tree.
    }
    \label{fig:lrf_grf_comparison}
\end{figure}

The simulated data set for the comparison consists of a piecewise linear model with changing slope in the first covariate of a $10$ dimensional data set. 
\begin{align}
    X_i &\sim N(0, I_{10})\\
    Y_i &= 3 X_i[1] \mathbf{1}_{(X_i[1] > 0) }  + (- 3 X_i[1]) \mathbf{1}_{(X_i[1] \leq 0) }
\end{align}
In this simulated example, we generate $n=500$ points from the above scheme, and fit a Linear Random Forest, with one tree, and a Local Linear Forest, with 11 trees on the resulting data \footnote{We use 11 trees for the Local Linear Forest as the software implementation does not allow for predictions with ten or less trees}.
Figure \ref{fig:lrf_grf_comparison} displays the difference in fits between Linear Random Forest and Local Linear Forest in this example. 

\section{Splitting on a Categorical Feature}
\label{appendix:categorical_split}
To split on a categorical feature, we use one-hot encoding to split based on equal or not equal to the given feature category. In order to evaluate the split RSS, we examine linear models fit on the set of observations containing the feature and the set not containing the feature. 
In order to evaluate this split quickly, we make use of the fact that we can quickly compute RSS components by keeping track of the total aggregated sum of outer products. 
\begin{center}
    Let $\ G_{\mbox{\footnotesize Total}} = \sum_{i = 1}^n \XiO \XiOT$ \\\
    
    $G_{\mbox{\footnotesize LeftInitial}} = \begin{bmatrix} X_1 \\ 1\end{bmatrix} \begin{bmatrix} X_1^T & 1\end{bmatrix}$ \\\
    
    $G_{\mbox{\footnotesize RightInitial}} = \sum_{i = 2}^n \XiO \XiOT = G_{\mbox{\footnotesize Total}} - G_{\mbox{\footnotesize LeftInitial}}$ \\\
\end{center}

This means that given $G_{\mbox{\footnotesize Total}}$, we can immediately begin evaluating splits along any split feature by using the sequence of indices corresponding to the ascending sequence of values in the current split feature. 

This fact helps us quickly evaluate the sum of RSS of separate regressions on the two sides of inclusion/exclusion splits on categorical variables.

On a split on categorical variable $X(l)$, when evaluating the split RSS on the value of category k, the RSS components can be calculated as follows:

\begin{center}
            Let $E({k}) = \{i : X_{i}(l) = k\}$ \\\
            
    		$G_{\mbox{\footnotesize Left}} = \sum_{i \in E({k})} \XiO \XiOT \in \mathbb{R}^{d\times d}$\\\
    		
    		$G_{\mbox{\footnotesize Right}} = \sum_{i \notin E({k})} \XiO \XiOT \in \mathbb{R}^{d\times d}$\\\
    		            
    		$G_{\mbox{\footnotesize Right}} = G_{\mbox{\footnotesize Total}} - G_{\mbox{\footnotesize Left}}$\\\

\end{center}

The same method can be used to update $S_{Left}$ and $S_{Right}$ at each step, so we can use this update rule generally to quickly compute the RSS for categorical splits in the algorithm that follows.

\begin{algorithm}[H] 
        \hspace*{\algorithmicindent} \textbf{Input:}
		Features: $X \in \mathbb{R}^{n\times d}$,
		\inputNL Dependent Outcome: $Y \in \mathbb{R}^{n}$,
		\inputNL overfitPenalty (regularization for split): $\lambda \in \mathbb{R}^+$,\\
		\hspace*{\algorithmicindent} \textbf{Output:} 
		Best Split point k
		\begin{algorithmic}[1]
			\Procedure{FindBestSplitRidgeCategorical}{}
			\item[]
			\item[]
			Initialization:
			\State $\ G_{Total} = \sum_{i = 1}^n \XiO \XiOT$
						
			\State $S_{Total} \leftarrow \sum_{i = 1}^n Y_i \XiO$
			
			\For{$category\_k = 1, \ldots, l$}
			\State $E({k}) \leftarrow \{i : X_{i}(splitFeat) = category\_k\}$
			
			\State $G_{L} \leftarrow \sum_{i \in E({k})} \XiO\XiOT$
			
			\State $S_{L} \leftarrow \sum_{i \in E({k})} Y_i\XiO$
			
			\State $G_{R} \leftarrow G_{Total} - G_{L}$
			
			\State $S_{R} \leftarrow S_{Total} - S_{L}$
			
			\State $A^{-1}_{R} \leftarrow \left(G_R + \lambda \begin{bmatrix}\mathcal{I}_d & 0 \\0 & 0\end{bmatrix}\right)^{-1}$
			\State $A^{-1}_{L} \leftarrow \left(G_L + \lambda \begin{bmatrix}\mathcal{I}_d & 0 \\0 & 0\end{bmatrix}\right)^{-1}$

			\State Compute the RSS sum for the current split:
			\begin{align*}
			\mbox{RSS}_k 
			\leftarrow 
				&S_{L}^T A^{-1}_{L} G_{L} A^{-1}_{L} S_{L} - 2 S_{L}^T A_{L}^{-1} S_{L})
				  \\+~ 
				&S_{R}^T A^{-1}_{R} G_{R}) A^{-1}_{R} S_{R} - 2 S_{R}^T A_{R}^{-1}
				 S_{R}.
			\end{align*}
			\EndFor
			\State \Return $(argmin_{(k)} \mbox{RSS}_k)$
			\EndProcedure
		\end{algorithmic}
		\caption{Find Best Split for Categorical Features}
		\label{algo:RSSonlineCategorical}	
\end{algorithm}

\subsection*{Runtime Analysis of Finding Categorical Split Point}

Using a set to keep track of category membership, we can create the set in $O(NlogN)$ time and access a member of any specific category in amortized constant time. 
Once we begin iterating through the categories, we can access the elements and create the left model RSS components in $O(|K|)$ where $|K|$ is the size of category k, and using the $G_{total}$ and $S_{total}$ matrices, we can find the right model RSS components in $O(d^2)$ time once we calculate $G_{L}$ and $S_{L}$. As the sum of sizes of the various categories sums to the number of observations, we end up doing the same number of RSS component update steps as the continuous case as well as one additional step to get the right RSS components for each category. The overall asymptotic runtime remains $O(NlogN + Nd^2)$.
\newpage

\section{Splitting Algorithm and Runtime}
\label{Appendix_runtime_and_algo}
For a given $\lambda > 0$ and leaf, $L$, the ridge-regression \citep{hoerl1970ridge} is defined as $(\hat \beta, \hat c)$ that minimize the ridge penalized least squares equation, 
\begin{equation}
	(\hat \beta, \hat c) = \argmin_{(\beta, c)}
	\sum_{i : X_i \in L} \left(Y_i - X_i^T \beta  - c\right)^2 +\lambda \|\beta\|_2^2,
\end{equation}
and a closed form solution of this problem is given by:
\begin{align}
		&\begin{bmatrix}\hat \beta \\ \hat c\end{bmatrix} 
		=
		\left({\sum_{i = 1}^n 
		\begin{bmatrix}X_i \\1 \end{bmatrix}
		\begin{bmatrix}X_i^T&1\end{bmatrix} 
		+
		\lambda 
		\begin{bmatrix} 
			\mathcal{I}_d & 0 \\
			0 & 0
		\end{bmatrix}}\right)^{-1}
		\sum_{i = 1}^n Y_i \begin{bmatrix} X_i \\ 1 \end{bmatrix}.
\end{align}	
Recall that for a set $H \subset \R^d$, we define
\begin{align*}
     A_H&\define \sum_{i : X_i \in H}
				\begin{bmatrix}X_i \\1 \end{bmatrix}
				\begin{bmatrix}X_i^T&1\end{bmatrix} 
				+
				\lambda 
				\begin{bmatrix} 
				\mathcal{I}_d & 0 \\
				0 & 0
				\end{bmatrix}\\
	S_H &\define \sum_{i : X_i \in H}Y_i \XiO,\\
	G_H &\define \sum_{i : X_i \in H}\XiO \XiOT.
\end{align*}
With this notation, we can decompose $\RSS(k)$:
\begin{align*}
\RSS(k)
&= 
\sum_{i : X_i \in L(k)}\left(Y_i - \XiOT A_{L(k)}^{-1} S_{L(k)}^{\phantom{-}}\right)^2
\\&+
\sum_{i : X_i \in R(k)}\left(Y_i - \XiOT {A_{R(k)}^{-1}} S_{R(k)}^{\phantom{-}}\right)^2 \\
&=  
\Phi_{L(k)} + \Phi_{R(k)} + \sum_{i = 1}^n Y_i ^2.
\end{align*}
Here, we use the definition which gives $\Phi_{H}$ for some $H \subset \R^d$ by:
\begin{equation} \label{eq:PhiDecomp}
\Phi_H \define S_{H}^T A_{H}^{-1} 
G_{H}^{\phantom{-}}
A_{H}^{-1} S_{H}^{\phantom{-}}
- 2 S_{H}^T A_{H}^{-1} S_{H}^{\phantom{-}}.
\end{equation}
As $\sum_{i = 1}^n Y_i ^2 $ is constant in $k$, it can be discarded when considering the optimization problem and thus:
\begin{equation} \label{eq:RSSequivalent}
\argmin_k~\RSS(k)
= 
\argmin_k ~ \Phi_L(k) + \Phi_R(k).
\end{equation}

\subsection{Pseudo Code of the Splitting Algorithm}
In order to have a manageable overall runtime, the algorithm needs to quickly find the minimizer of (\ref{eq:minimizing_for_k}). The fundamental idea behind the fast splitting algorithm is that we reuse many terms that were computed when finding $\mbox{RSS}(k)$ in order to calculate $\mbox{RSS}(k+1)$. This enables us to loop very quickly from $k=1$ to $k=n$ and calculate the RSS for each iteration.
Specifically, the algorithm can be expressed in the following pseudo code.
\begin{enumerate}
	\item Compute $S_{L(1)}$, $A_{L(1)}^{-1}$, $G_{L(1)}$, $S_{R(1)}$, $A_{R(1)}^{-1}$, $G_{R(1)}$, and $\mbox{RSS}(1)$.
	\item Compute the $\mbox{RSS}(k)$ for $k \ge 2$ in an iterative way: 
	\begin{enumerate}
		\item $S_{L(k)}$, $G_{L(k)}$, $S_{R(k)}$,  and $G_{R(k)}$ can be directly computed from $S_{L(k - 1)}$, $G_{L(k - 1)}$, $S_{R(k - 1)}$,  and $G_{R(k - 1)}$ by a simple addition or subtraction. 
		\item For $A_{L(k)}^{-1}$ and $A_{R(k)}^{-1}$, we use the Sherman-Morrison Formula:
			\begin{align} 
			A_{L(k)}^{-1} &= 
			A_{L(k - 1)}^{-1} - 
			\frac{
				A_{L(k - 1)}^{-1}\begin{bmatrix}x_{k} \\1\end{bmatrix}\begin{bmatrix}x_{k}^T &1\end{bmatrix}A_{L(k - 1)}^{-1}
			}{
				1+\begin{bmatrix}x_{k}^T &1\end{bmatrix}A_{L(k - 1)}^{-1}\begin{bmatrix}x_{k} \\1\end{bmatrix}
			},
			\\
			A_{R(k)}^{-1} &= 
			A_{R(k - 1)}^{-1} + 
			\frac{
				A_{R(k - 1)}^{-1}\begin{bmatrix}x_{k} \\1\end{bmatrix}\begin{bmatrix}x_{k}^T &1\end{bmatrix}A_{R(k - 1)}^{-1}
			}{
				1-\begin{bmatrix}x_{k}^T &1\end{bmatrix}A_{R(k - 1)}^{-1}\begin{bmatrix}x_{k} \\1\end{bmatrix}
			}.
			\end{align} 
	\end{enumerate}
\end{enumerate}
An explicit implementation of the split algorithm can be found in Algorithm \ref{algo:RSSonlineContinuous}.

\begin{algorithm}[H]
		\hspace*{\algorithmicindent} \textbf{Input:}
		Features: $X \in \mathbb{R}^{n\times d}$,
		\inputNL Dependent outcome: $Y \in \mathbb{R}^{n}$,
		\inputNL overfitPenalty (regularization for split): $\lambda \in \mathbb{R}^+$,\\
		\hspace*{\algorithmicindent} \textbf{Output:} 
		Best Split point k
		\begin{algorithmic}[1]
		\Procedure{FindBestSplitRidge}{}
			\item[]
			\item[]
			Initialization:
			\State $A^{-1}_{L1} \leftarrow \left(\begin{bmatrix} X_1 \\ 1\end{bmatrix} \begin{bmatrix} X_1^T & 1\end{bmatrix} + \lambda\begin{bmatrix}\mathcal{I}_d & 0 \\0 & 0\end{bmatrix}\right)^{-1}$
			\State $A^{-1}_{R1} \leftarrow \left(\sum_{i = 2}^{n} \XiO \XiOT + \lambda \begin{bmatrix}\mathcal{I}_d & 0 \\0 & 0\end{bmatrix}\right)^{-1}$
			\State $S_{L1} \leftarrow Y_1 \begin{bmatrix} X_1 \\ 1\end{bmatrix}$
			\State $S_{R1} \leftarrow \sum_{i = 2}^n Y_i \XiO$
			
			\State $G_{L1} \leftarrow  \begin{bmatrix} X_1 \\ 1\end{bmatrix} \begin{bmatrix} X_1^T & 1\end{bmatrix}$
			\State $G_{R1} \leftarrow \sum_{i = 2}^n \XiO \XiOT$
			
			\State Compute the RSS sum:
			\begin{align*}
				\mbox{RSS}_1 
				\leftarrow 
				&S_{L1}^T A^{-1}_{L1} G_{L1} A^{-1}_{L1} S_{L1} - 2 S_{L1}^T A_{L1}^{-1} S_{L1}
				\\+~ 
				&S_{R1}^T A^{-1}_{R1} G_{R1} A^{-1}_{R1} S_{R1} - 2 S_{R1}^T A_{R1}^{-1}
				S_{R1}.
			\end{align*}

			\For{$k = 2, \ldots, n$}
			\State $A^{-1}_{L(k)} \leftarrow \text{Update\_A\_inv}(A^{-1}_{L(k-1)}, X_k, \text{leftNode} = TRUE)$
			\State $A^{-1}_{R(k)} \leftarrow \text{Update\_A\_inv}(A^{-1}_{R(k-1)}, X_k, \text{leftNode} = FALSE)$

			\State $S_{L(k)} \leftarrow S_{L(k-1)} + Y_k \XkO$
			\State $S_{R(k)} \leftarrow S_{R(k-1)} - Y_k \XkO$

			\State $G_{L(k)} \leftarrow G_{L(k-1)} + \XkO \XkOT$
			\State $G_{R(k)} \leftarrow G_{R(k-1)} - \XkO \XkOT$

			\State Compute the RSS sum for the current split:
			\begin{align*}
			\mbox{RSS}_k 
			\leftarrow 
				&S_{L(k)}^T A^{-1}_{L(k)} G_{L(k)} A^{-1}_{L(k)} S_{L(k)} - 2 S_{L(k)}^T A_{L(k)}^{-1} S_{L(k)})
				  \\+~ 
				&S_{R(k)}^T A^{-1}_{R(k)} G_{R(k)}) A^{-1}_{R(k)} S_{R(k)} - 2 S_{R(k)}^T A_{R(k)}^{-1}
				 S_{R(k)}.
			\end{align*}
			\EndFor
			\State \Return $(\argmin_{k} \mbox{RSS}_k)$
			\EndProcedure
		\end{algorithmic}
		\caption{Find Split to Minimize Sum of RSS}
		\label{algo:RSSonlineContinuous}	
		\algcomment{Update\_A\_inv is defined in Algorithm \ref{algo:updateA}}
	\end{algorithm}

\begin{algorithm}[H]
		\hspace*{\algorithmicindent} \textbf{Input:} $A^{-1}_{k-1}\in \mathbb{R}^{(d+1)\times (d+1)}$, 
		\inputNL $X_k \in\mathbb{R}^d$, 
		\inputNL leftNode (indicator whether this updates $A_{L(k)}^{-1}$ or $A_{R(k)}^{-1}$)\\
		\hspace*{\algorithmicindent} \textbf{Output:} 
		Updated matrix $A^{-1}_{k}$
		\begin{algorithmic}[1]
			\Procedure{Update\_A\_inv}{}
			\State 
			$z_{k}  \leftarrow A_{k-1}^{-1}  \begin{bmatrix}X_{k}\\1\end{bmatrix}$
			\If{leftNode}
				\State
				$g_{k} \leftarrow\dfrac{-z_{k}z_{k}^{T}}{1+\begin{bmatrix}X_{k}^{T} & 1 \end{bmatrix} z_{k}}$
			\Else
				\State
				$g_{k} \leftarrow\dfrac{z_{k}z_{k}^{T}}{1-\begin{bmatrix}X_{k}^{T} & 1 \end{bmatrix} z_{k}}$
			\EndIf 
			\State \Return $A_{k-1}^{-1} + g_{k}$ 
			\EndProcedure
		\end{algorithmic}
		\caption{Update the $A^{-1}$ Component of the RSS}
		\label{algo:updateA}
\end{algorithm}

\subsection{Runtime Analysis of Finding Split Points}
\label{runtimeProof}
The ability to use an online update for calculating the iterated RSS at each step is crucial for maintaining a quasilinear runtime. 
Here we will provide a detailed breakdown of the runtime for calculating the best split point on a given feature.
As we have several steps for updating the RSS components, the runtime depends on both the number of observations, as well as the number of features and therefore may be affected by either. 
We begin each split by sorting the current split feature taking $\O(n \log n)$ time. Within the split iterations, we iterate over the entire range of split values once, however, at each step we must update the RSS components. 

While updating the $A^{-1}$ component, as we use the Sherman-Morrison Formula to update the inverse of the sum with an outer product, we must compute one matrix vector product ($\O(d^2)$), one vector outer product ($\O(d^2)$), one vector inner product ($\O(d^2)$), division of a matrix by scalars and addition of matrices (both $\O(d^2)$). 
While updating the $G$ component, we need to both add and subtract an outer product (both $\O(d^2)$), and while updating the $S$ component, we need to add and subtract a vector ($\O(d)$).
At each step of the iteration, we must evaluate the RSS of the right and left models. To do this, we need 8 matrix vector products, each of which is $\O(d^2)$, and 4 vector inner products, each of which is $\O(d)$.
Putting these parts together gives us a total runtime of $\O(n \log n + n d^2)$ to find the best split point for a given node with $n$ observations and a $d$-dimensional feature space. 

\subsection{Early Stopping} \label{sec:earlystopping}
As we will see in Section \ref{section:simulationstudies}, early stopping can prevent overfitting in the regression tree algorithm and the RF algorithm. 
Furthermore, as we discuss in Section \ref{section:interpretability}, it leads to well performing yet shallow trees that are much easier to understand and interpret. 

We use a one step look-ahead stopping criteria to stop the trees from growing too deep. Specifically, we first compute a potential split as outlined in the algorithm above. 
We then use cross validation to compute the $R^2$ increase of this split and only accept it, if the increase of the split is larger than the specified \code{minSplitGain} parameter.
A larger \code{minSplitGain} thus leads to smaller trees.
The precise description of this parameter can be found in Algorithm \ref{algo:EarlyStopping}.

\begin{algorithm}
		\hspace*{\algorithmicindent} \textbf{Input:}
		Features: $X \in \mathbb{R}^{n\times d}$,
		\inputNL Dependent outcome: $Y \in \mathbb{R}^{n}$,
		\inputNL Indices of potential left child: $L \subset \{1, \ldots, n\}$,
		\inputNL minSplitGain: $m \in \R$,
		\inputNL numTimesCV: $k \in \{2, \ldots, n\}$,
		\inputNL overfitPenalty: $\lambda > 0$
		\\
		\hspace*{\algorithmicindent} \textbf{Output:} 
		Boolean: TRUE if split is accepted, FALSE otherwise
		\begin{algorithmic}[1]
			\Procedure{Check\_split}{}
			\State Partition $\{1, \ldots, n\}$ into $k$ disjoint subsets: $\{S_1, \ldots, S_k\}$.
			\For{$i$ in $\{1, \ldots, k\}$}
			    \item[]
			    \item[] ~~~~~~~~~Predict the outcome without the split:
			    \State For $j \in S_i$ set
			    \begin{align*}
            		\hat Y^p_j 
            		=
        		    \begin{bmatrix}X_j \\1 \end{bmatrix}
            		\left({\sum_{k \in \bar S_i} 
            		\begin{bmatrix}X_k \\1 \end{bmatrix}
            		\begin{bmatrix}X_k^T&1\end{bmatrix} 
            		+
            		\lambda 
            		\begin{bmatrix} 
            			\mathcal{I}_d & 0 \\
            			0 & 0
            		\end{bmatrix}}\right)^{-1}
            		\sum_{k \in \bar S_i} Y_k \begin{bmatrix} X_k \\ 1 \end{bmatrix}.
            	\end{align*}
            \item[]
			\item[] ~~~~~~~~~Predict the outcome with the split:
            	\State
			    For $j \in S_i \cap L$ set
			    \begin{align*}
            		\hat Y^c_j 
            		=
        		    \begin{bmatrix}X_j \\1 \end{bmatrix}
            		\left({\sum_{k \in \bar S_i\cap L} 
            		\begin{bmatrix}X_k \\1 \end{bmatrix}
            		\begin{bmatrix}X_k^T&1\end{bmatrix} 
            		+
            		\lambda 
            		\begin{bmatrix} 
            			\mathcal{I}_d & 0 \\
            			0 & 0
            		\end{bmatrix}}\right)^{-1}
            		\sum_{k \in \bar S_i\cap L} Y_k \begin{bmatrix} X_k \\ 1 \end{bmatrix}.
            	\end{align*}
			    
			    \State
			    For $j \in S_i \cap \bar L$ set
			    \begin{align*}
            		\hat Y^c_j 
            		=
        		    \begin{bmatrix}X_j \\1 \end{bmatrix}
            		\left({\sum_{k \in \bar S_i\cap \bar L} 
            		\begin{bmatrix}X_k \\1 \end{bmatrix}
            		\begin{bmatrix}X_k^T&1\end{bmatrix} 
            		+
            		\lambda 
            		\begin{bmatrix} 
            			\mathcal{I}_d & 0 \\
            			0 & 0
            		\end{bmatrix}}\right)^{-1}
            		\sum_{k \in \bar S_i\cap \bar L} Y_k \begin{bmatrix} X_k \\ 1 \end{bmatrix}.
            	\end{align*}
            	
			\EndFor
			\State Compute the estimated RSS with and without split:
			\begin{align*}
			\mbox{RSS}^{c}  = \sum_{i = 1}^n \Big(Y_i - \hat Y^c_i \Big)^2 &&\mbox{and}&&
			\mbox{RSS}^{p} =  \sum_{i = 1}^n \Big(Y_i - \hat Y^p_i \Big)^2
			\end{align*}
			\State 
			Compute the total variation: 
			$
			\mbox{tV} = \sum_{i = 1}^n \left(Y_i - \bar Y\right)^2
			$
			\If{$(\mbox{RSS}^c - \mbox{RSS}^p)/\mbox{tV} > m$}
			
				\Return TRUE
				
			\Else
			
				\Return FALSE 
				
			\EndIf 
			\EndProcedure
		\end{algorithmic}
		\caption{Early Stopping: The One Step Look-Ahead Algorithm}
		\label{algo:EarlyStopping}
		\algcomment{For a set $S$ we define its compliment as $\bar S := \{i : 1\le i \le n\mbox{ and } i \not \in S \}$.}
\end{algorithm}

\section{Tuned Simulation Hyperparameters}
\label{appendix:hyperparams}
\begin{table}[H]
\begin{center}
\begin{tabular}{ |c|c|c|c| } 
\hline
\textbf{Estimator} & \textbf{Package} & \textbf{Tuned Hyperparameters} \\
\hline
Random Forest & \ranger & mtry, min.node.size, num.trees, sample.fraction \\ 
\hline
Random Forest & \forestry & mtry, nodesizeStrictSpl, ntree, sample.fraction \\ 
\hline
Linear Random Forest & \forestry & mtry, nodesizeStrictSpl, ntree, \\ 
& & sample.fraction, overfitPenalty, minSplitGain \\ 
\hline
Regularized Linear Model & \glmnet & alpha, lambda \\ 
\hline
Cubist & \Cubist & committees, extrapolation, neighbors \\ 
\hline
Bayesian Additive Regression Trees & \dbarts & base, power, ntree, sigdf \\ 
\hline
Local Linear Forest & \grf &  mtry, min.node.size, num.trees, sample.fraction \\ 
\hline
Linear CART & \forestry &  mtry, nodesizeStrictSpl, \\
& & overfitPenalty, minSplitGain \\ 
\hline
\end{tabular}
\caption{\label{tab:hyperparameters_tuned} Above displays the software packages and tuned hyperparameters used by the \caret package for each estimator in Section \ref{section:simulationstudies}. For the exact hyperparameter values selected by the \caret package, see the replication archive at \textbf{https://github.com/forestry-labs/RidgePaperReplication}.}
\end{center}
\end{table}

\newpage
\section{Generating Random Step Function}\label{sec:AppCreatestepResSurf}

\textbf{The Simulated-Step-Function in Section \ref{section:simulationstudies} was generated according to the following scheme:}

\begin{algorithm}
		\hspace*{\algorithmicindent} \textbf{Input:} numLevels (number of random levels for step function),
		\inputNL n (dimension of data)\\
		\hspace*{\algorithmicindent} \textbf{Output:} Independent Input: $X \in \mathbb{R}^{n\times d}$,
		\inputNL Step Function Outcome: $Y_{step} \in \mathbb{R}^{n}$,
		\begin{algorithmic}[1]
			\Procedure{Generate\_Simulated\_Step}{}
			\State 
			$X_{i=1}^{n}  \leftarrow Normal(0,1)^{10}$\\
			\State
			$Y_{i=1}^{numLevels}  \leftarrow Unif(-10,10)$\\
			\State 
			$X_{sample} \leftarrow X[sample(1:nrow(X),numLevels, replace = FALSE),]$\\
		    \State
		    $f_s \leftarrow forestry(x = X_{sample}, y = Y, nodeSize = 1)$\\
			\State 
			$Y_{step} \leftarrow predict(f_s, feature.new = X)$\\
			\State
			\Return $(X,Y_{step})$ 
			\EndProcedure
		\end{algorithmic}
		\caption{Generate Simulated Step}
		\label{algo:generateStepFunction}
\end{algorithm}

\newpage

\section{Optimal Regularization Parameters}
\label{appendix:optimal_lambda}

As noted in Section \ref{section:simulationstudies}, the value of $\lambda$ used for the ridge regression can allow linear random forests to behave like either a regularized linear model or a standard random forest.

In order to take advantage of this adaptivity, we allow the user to specify the desired regularization parameter in our software using the \textit{overfitPenalty} parameter. 
In Section \ref{section:simulationstudies}, as we have a variety of data sets which exhibit varying smoothness, we select the optimal \textit{overfitPenalty} using the \textbf{caret} package \citep{kuhn2008building}.

As the value of $\lambda$ is considered in the optimization problem which finds the optimal splitting point at each iteration, we study only the case where we have a globally fixed \textit{overfitPenalty}.
It is also possible to replace the final value of $\lambda$ in each leaf node with a value different from that used to select the splits for the tree. 
As there may be many leaf nodes which may make complicated estimation schemes for individual $\lambda$'s computationally intensive, we suggest selecting $\lambda$ for each node following a simple rule of thumb.
One such rule, inspired by \citep{donoho1995}, would select $\lambda$ for each node based on:
\begin{align}
    \lambda \ \alpha \ \sigma \sqrt{\frac{\log{p}}{n}}
\end{align}
Where $\sigma$ is the square root of the variance of outcomes in the leaf node, $p$ is the number of linear features the ridge regression is fit on, and $n$ is the number of observations in the node.

\newpage 

\section{Software Solutions}
\label{appendix:software}

To aid researchers in analyzing regression trees, we implemented a plotting function in the \forestry package that is fast, easy to use, and flexible.\footnote{\forestry is available at \url{https://github.com/forestry-labs/Rforestry}.} It enables researchers to quickly look at many trees in a RF algorithm.
The analysis done in Section \ref{sec:SL_interpretability} was created using this package and using the \code{tree.id} parameter to specify different trees to plot. 

\begin{lstlisting}
rft <- forestry(x = x,
                y = y, 
                minSplitGain = .005,
                linear = TRUE,
                overfitPenalty = 1e-8,
                linFeats = 1:4,
                splitratio = .5)
plot(forest, tree.id = 1)
\end{lstlisting}
In the above code segment, the \code{linear} parameter specifies to use the ridge splitting criterion, the \code{linFeats} parameter specifies the columns of the design matrix on which to evaluate the ridge split, and the \code{overfitPenalty} parameter gives the ridge penalty for the splitting algorithm- as detailed in Algorithm \ref{algo:RSSonlineContinuous}.
The \code{minSplitGain} parameter specifies the percentage increase in $R^2$ required for any split to be accepted.
We recommend using a smaller \code{minSplitGain} to draw even more personalized conclusions.
For details on the role of \code{minSplitGain} in the acceptance of a split, refer to Algorithm \ref{algo:EarlyStopping}.

\newpage

\section{Interpretability}
\label{appendix:interpretability}

\subsection{Voter Turnout Data Set}

We use a data set from a large field experiment that has been conducted by \cite{GerberGreenLarimer} to measure the effectiveness of four different mailers on voter turnout.
The data set contains 344,084 potential voters in the August 2006 Michigan statewide election with a wide range of offices and proposals on the ballot.
The sample was restricted to voters that voted in the 2004 general election, because the turnout in the 2006 election was expected to be extremely low among potential voters who did not vote in the 2004 general election. 

The specific mailers can be found in \cite{GerberGreenLarimer} and we briefly describe the mailers as follows. 
Each mailer carries the message "DO YOUR CIVIC DUTY --- VOTE!" and applies a different level of social pressure. We present the different mailers ordered by the amount of social pressure they put on the recipients. 
\begin{itemize}
    \item Civic Duty (CD): The Civic Duty mailer contains the least amount of social pressure and only reminds the recipient that {the whole point of democracy is that citizens are active participants in government}. 
    \item Hawthorne (HT): Households receiving the Hawthorne mailer were told "YOU ARE BEING STUDIED!" and it explains that voter turnout will be studied in the upcoming August primary election, but whether or not an individual votes "will remain confidential and will not be disclosed to anyone else."
    \item Self (SE):
    The Self mailer exerts even more pressure. It informs the recipient that \say{WHO VOTES IS PUBLIC INFORMATION!} and presents a table showing who voted in the past two elections within the household to which the letter was addressed. 
    It also contains a field for the upcoming August 2006 election, and it promises that an updated table will be sent to households after the election. 
    \item Neighbors (NE):
    The Neighbors mailer increases the social pressure even more and starts with the rhetorical question: \say{WHAT IF YOUR NEIGHBORS KNEW WHETHER YOU VOTED?} 
    It lists the voting records not only of the people living in the household, but also of those people living close by and it promises to send an updated chart after the election. This treatment mailer implies that the voting turnout of the recipients are publicised to their neighbors and thereby creates the strongest social pressure. 
\end{itemize}

The randomization was done on a household level. The set of all households was split into  four treatment groups and a control group. Each of the four treatment groups contained about 11\% of all voters, while the control group contained about 56\% of all households.

\subsection{S-Learner Using LRF}

While there are several ways in which the S-Learner could be implemented using LRF, we believe that the way we have done so offers several advantages.

Our implementation utilizes the linear model in each node in order to handle several treatments simultaneously. 
Thus for a leaf, $L$, the regression coefficients for each treatment indicator becomes estimates for the average treatment effect for that treatment for observations in that leaf,
      \begin{equation}
    	(\hat \tau, \hat b) = \argmin_{(\tau, b)}
    	\sum_{i : X_i \in L} \left(Y_i - W_i^T \tau  - b\right)^2 +\lambda \|\tau\|_2^2.
    \end{equation}
    Where $W_i \in \{0,1\}^4$ is the four dimensional vector containing the indicator for the treatment group of unit $i$ and  
    $X_i$ and $Y_i$ are the features and the dependent outcome of unit $i$. 
As the tree splits are done on the other covariates, this allows the tree structure to uncover heterogeneity in the CATE function while providing estimates for the CATE of all four treatments.
The use of many trees, with bootstrap samples to build the trees on, allows us to reduce the variance of our CATE estimates. 

Due to the use of \textit{honesty} when constructing each tree, the linear model in each node is fit on a data set which is independent of the data set used to construct the tree. 

Finally, as we make use of the stopping criterion (Algorithm \ref{algo:EarlyStopping}), we prevent splits which do not sufficiently uncover heterogeneity of the treatment effects, which prevents the trees from overfitting, and provides much more interpretable estimators.

\subsection{Additional Trees}

Below we plot several additional trees from the ensemble studied in Section \ref{section:interpretability}.
In Section \ref{section:interpretability} we focus on conclusions drawn from a single tree of the ensemble, however we can see that these conclusions are supported as we examine additional trees in the forest. 

We still see a subgroup of voters where there seems to be very little interest in participating in past or future elections, regardless of treatment received.
We also see a subgroup which seems to be highly engaged in voting, however still sees some of the highest treatment effects for the various mailers.
Finally, there are consistently splits on age in the early 50s, suggesting that this is a stable cutpoint which determines a large difference in potential voter turnout.

\begin{figure}[H]
	\begin{center}	
		\begin{subfigure}[b]{.895\textwidth}
			\centering
		\includegraphics[height=.68\linewidth]{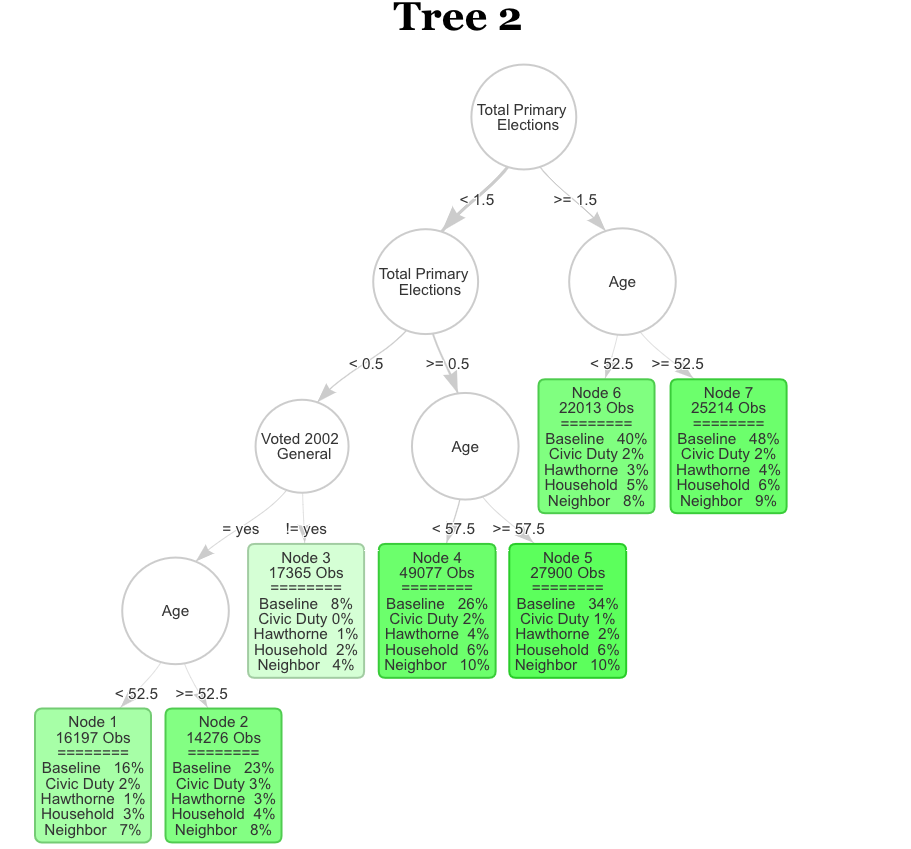}
		\caption{}
		\end{subfigure}
		\begin{subfigure}[b]{.895\textwidth}
			\centering
		\includegraphics[height=.68\linewidth]{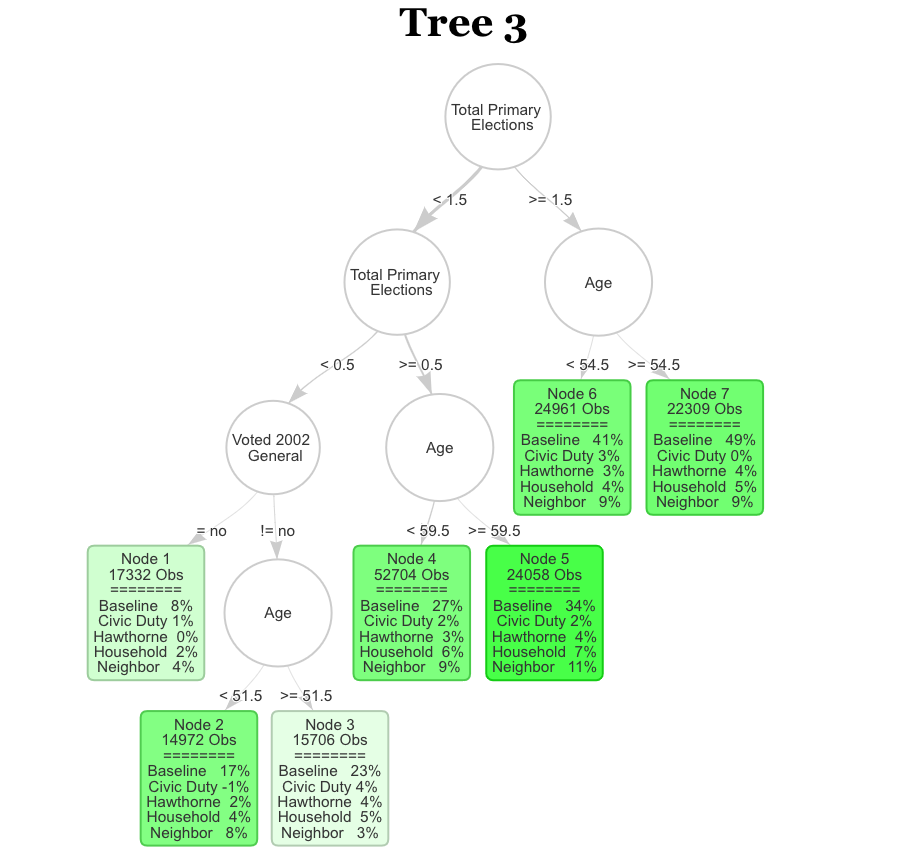}
		\caption{}
		\end{subfigure} 
	\end{center}
\end{figure}	
\begin{figure}[H]\ContinuedFloat
	\begin{center}
		\begin{subfigure}[b]{.895\textwidth}
			\centering
		\includegraphics[height=.78\linewidth]{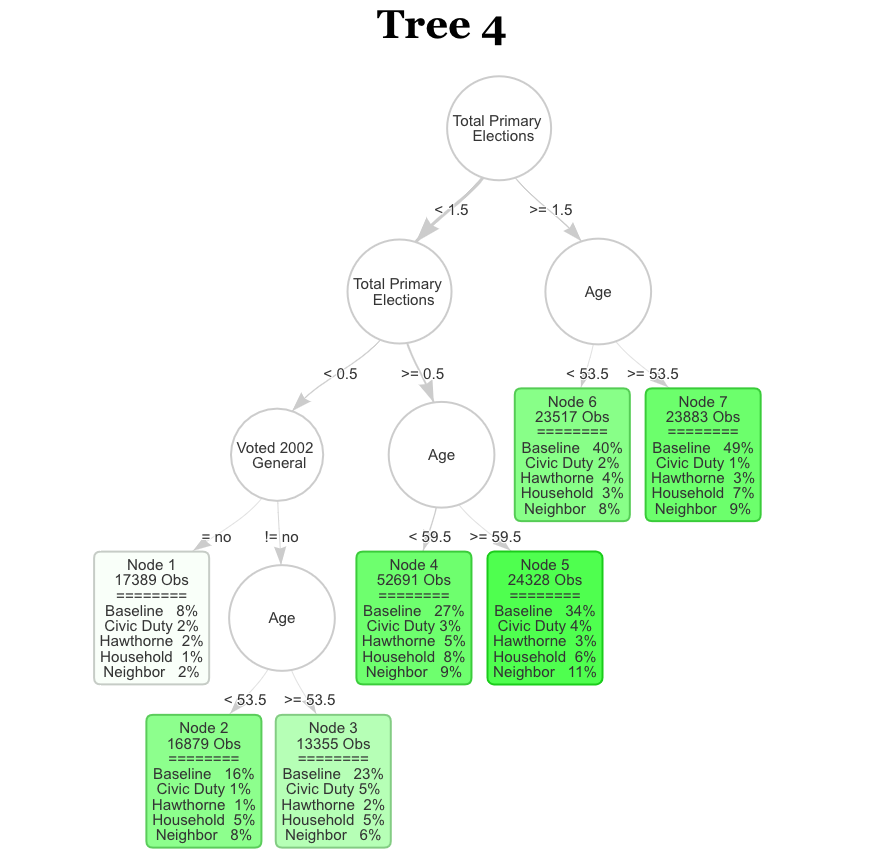}
		\caption{}
		\end{subfigure} 
	\end{center}
	\caption{The next three trees of the S-Learner as described in Section \ref{sec:SL_interpretability}. The first row in each leaf contains the number of observations in the averaging set that fall into the leaf. The second part of each leaf displays the regression coefficients. 
	\code{Baseline} stands for untreated base turnout of that leaf and it can be interpreted as the proportion of units that fall into that leaf who voted in the 2004 general election. 
    Each coefficient corresponds to the ATE of the specific mailer in the leaf.
	The color strength is chosen proportional to the treatment effect for the neighbors treatment.
	}
\end{figure}

\newpage

\section{Interpreting LRF Outputs}

\subsection{Local Linear Aggregation}

As the base learners in Linear Random Forests vary substantially from those in the standard random forest algorithm, it is of interest to see how the different functional form of the base learners effects the predictions of the ensemble.

As the predictions of an LRF ensemble come from a locally linear model, this gives information not only about the point predictions, but also how these predictions might change when the covariates are perturbed slightly.

To aid researchers in utilizing the smooth regression surface of linearly aggregated trees, we implemented a prediction function in the \forestry package that allows users to examine the forest wide average coefficient for each covariate used to predict a given observation.\footnote{\forestry is available at \url{https://github.com/forestry-labs/Rforestry}.}

\begin{lstlisting}
rft <- forestry(x = x,
                y = y, 
                linear = TRUE,
                overfitPenalty = .1)
                
predict(rft, newdata = x, aggregation = "coefs")
\end{lstlisting}

One example of where this might be useful is in the estimation of incremental causal effects \citep{rothen2020incremental}.
A setting where practitioners might be interested in the change in treatment effect due to a slight perturbation of either a single or multiple covariates. 

\subsection{Variable Importance Metrics}

The use of linear aggregation functions also modifies how the standard random forest variable importance \citep{breiman2001random} behaves.
In order to understand the differences produced in the variable importances of different features, we show a simple simulation.

\subsubsection{Data Generating Process}

Again we follow the example from \cite{athey2019generalized} used in Section \ref{sec:high_dim_simulations} to study the performance of LRF in the prescence of many noisy covariates. 
We make one modification to this simulation in changing the dimension of the covariates from 100 to 10 and selecting a higher amount of noise.
The setup now becomes $X \sim N(0,1)^{10}$, then $Y$ is generated as follows:
\begin{align}
    Y = 2 \varsigma\left(X_{1}\right) \varsigma\left(X_{2}\right) + 2 \varsigma\left(X_{3}\right) + 3 \varsigma\left(X_{4}\right) + \epsilon
\end{align}
Where:
\begin{align}
    \varsigma(x)=\frac{2}{1+e^{-12(x-.5)}}
\end{align}
And $\epsilon \sim N(0,2)$.
As the true outcome depends on only the first four covariates, we wish to pick up these as the most important covariates.

\subsubsection{Results}

We train both a Linear Random Forest and the standard RF algorithm on the DGP above and present the normalized variable importances. 
For this simulation we take the standard RF variable importance introduced in \cite{breiman2001random}, and average our results over 100 Monte Carlo replications.

We can see that both LRF and RF are able to pick up the important features quite well in this example.
The normalized feature importances are nearly identical for both algorithms, however the unnormalized importances are consistently higher for the standard RF algorithm.
This suggests that the linear aggregation algorithm is less sensitive to perturbations in the covariates than the standard CART predictions.

\begin{figure}[H]
    \centering
    \includegraphics[width =.45\textwidth]{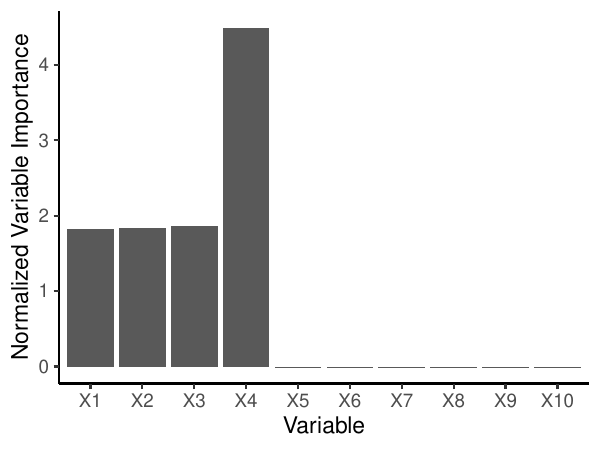}
    \includegraphics[width =.45\textwidth]{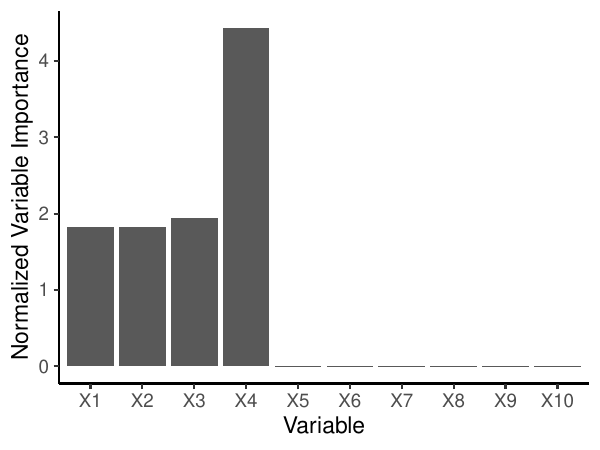}
    \caption{The left shows the mean normalized variable importances from LRF on simulated data, repeated over 100 Monte Carlo replications. The right shows the mean normalized variable importances from RF on simulated data, repeated over 100 Monte Carlo replications. The true outcome relies on only the first four covariates.}
\end{figure}

\newpage

\section{Runtime Comparison}
\label{appendix:runtime}

In order to show the practical implications of Theorem \ref{thm:runtime_result}, we show several simulations to show the magnitude of the speedup when running the LRF algorithm on real data sets. 
We generate data according to $X \sim N(0,1)^{p}$, with $Y$ generated according to:
\begin{align}
\label{runtime_sim}
    Y = .4 X_1 + 2 X_2 - .9 X_3 +.25 X_4 + \epsilon
\end{align}
Where $\epsilon \sim N(0,1)$.
We then generate data from this model for a range of dimensions $p$ and samples sizes $n$.

\begin{figure}[H]
    \centering
    \includegraphics[width =.45\textwidth]{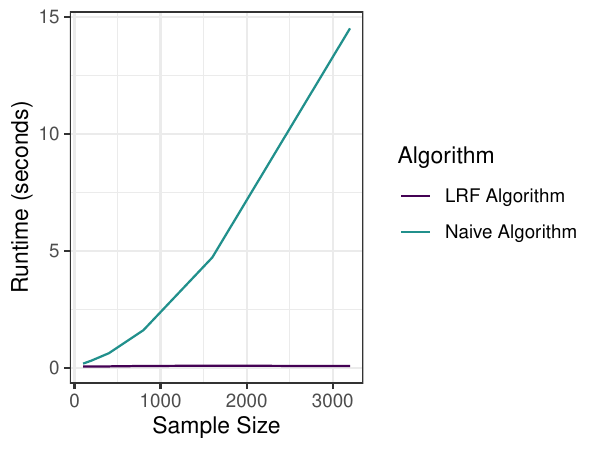}
    \includegraphics[width =.45\textwidth]{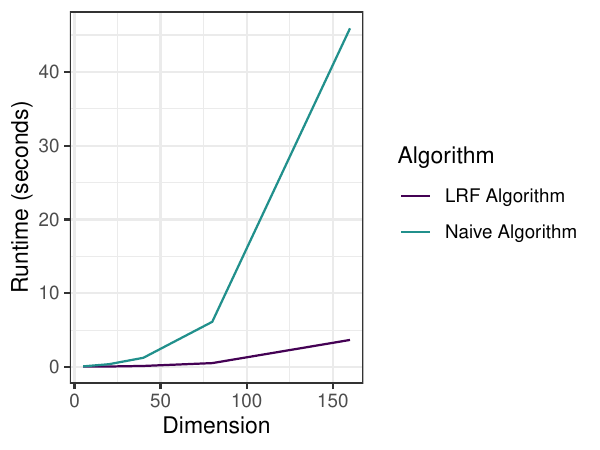}
    \caption{We run both the naive linear regression tree algorithm and the fast LRF algorithm from Section \ref{section:Algorithm} for 100 Monte Carlo replications on the same data generated according to Equation \ref{runtime_sim}. The left figure shows the mean timing of both algorithms as we keep the dimension fixed and vary the sample size, and the right figure shows the mean timing of both algorithms as we fix the sample size and vary the dimension.}
\end{figure}

\end{document}